\documentclass[12pt]{article}
\usepackage{ametsoc}
\linenumbers

\newcommand{\myabstract}{
   In this work we determine key model parameters for rapidly
   intensifying Hurricane Guillermo (1997) using the Ensemble Kalman
   Filter (EnKF). The approach is to utilize the EnKF as a tool to
   only estimate the parameter values of the model for a particular
   data set. The assimilation is performed using dual-Doppler radar
   observations obtained during the period of rapid intensification of
   Hurricane Guillermo. A unique aspect of Guillermo was that during
   the period of radar observations strong convective bursts,
   attributable to wind shear, formed primarily within the eastern
   semicircle of the eyewall. To reproduce this observed structure
   within a hurricane model, background wind shear of some magnitude
   must be specified; as well as turbulence and surface parameters
   appropriately specified so that the impact of the shear on the
   simulated hurricane vortex can be realized.  To identify the
   complex nonlinear interactions induced by changes in these
   parameters, an ensemble of model simulations have been conducted in
   which individual members were formulated by sampling the parameters
   within a certain range via a Latin hypercube approach. The ensemble
   and the data, derived latent heat and horizontal winds from the
   dual-Doppler radar observations, are utilized in the EnKF to obtain
   varying estimates of the model parameters. The parameters are
   estimated at each time instance, and a final parameter value is
   obtained by computing the average over time. Individual simulations
   were conducted using the estimates, with the simulation using
   latent heat parameter estimates producing the lowest overall model
   forecast error.
}
\begin{document}
%
\title{\textbf{\large{Determining Key Model Parameters of
Rapidly Intensifying Hurricane Guillermo(1997) using the Ensemble
Kalman Filter}}}
%
%
\author{\textsc{Humberto C. Godinez}\thanks{\textit{Corresponding author address:} 
				Humberto C. Godinez, Los Alamos
                                National Laboratory, MS B284,
				Los Alamos, NM 87545. 
                                \newline{E-mail: hgodinez@lanl.gov}} 
                                \textsc{ and Jon M. Reisner}
				\\
\textit{\footnotesize{Los Alamos National Laboratory, Los Alamos, New
Mexico}}
\and 
\centerline{\textsc{Alexandre O. Fierro}}\\
\centerline{\textit{\footnotesize{ Earth and Environmental Sciences
division/Space and Remote Sensing Group, Los Alamos National
Laboratory, Los Alamos, New Mexico}}}\\
\centerline{\textit{\footnotesize{ NOAA/Cooperative Institute for
Mesoscale Meteorological Studies, Norman, Oklahoma}}}
\and
\centerline{\textsc{Stephen R. Guimond}}\\
\centerline{\textit{\footnotesize{Center for Ocean-Atmospheric
Prediction Studies, Florida State University, Tallahassee, Florida}}}
\and
\centerline{\textsc{Jim Kao}}\\
\centerline{\textit{\footnotesize{Los Alamos National Laboratory, Los
Alamos, New Mexico}}}
}
%
\ifthenelse{\boolean{dc}}
{
\twocolumn[
\begin{@twocolumnfalse}
\amstitle

\begin{center}
\begin{minipage}{13.0cm}
\begin{abstract}
	\myabstract
	\newline
	\begin{center}
		\rule{38mm}{0.2mm}
	\end{center}
\end{abstract}
\end{minipage}
\end{center}
\end{@twocolumnfalse}
]
}
{
\amstitle
\begin{abstract}
\myabstract
\end{abstract}
}
\section{Introduction}
\label{sec:Introduction}

Hurricanes are among the most destructive and costliest natural forces
on Earth and hence it is important to improve the ability of numerical
models to forecast changes in their track, intensity, and structure.
But, accurate prediction depends on minimizing errors associated with
initial, environmental, and boundary conditions, numerical
formulations, and physical parameterizations. Though significant
progress has been made over the past decade with regard to using data
assimilation to primarily improve the initial state of a hurricane
(\cite{ZhangEtAl2009}; \cite{TornHakim2009}; \cite{ZoutEtAl2010}),
uncertainties still remain in other aspects of hurricane
prediction. 

\begin{figure}
   \centering
   \includegraphics[width=\linewidth]{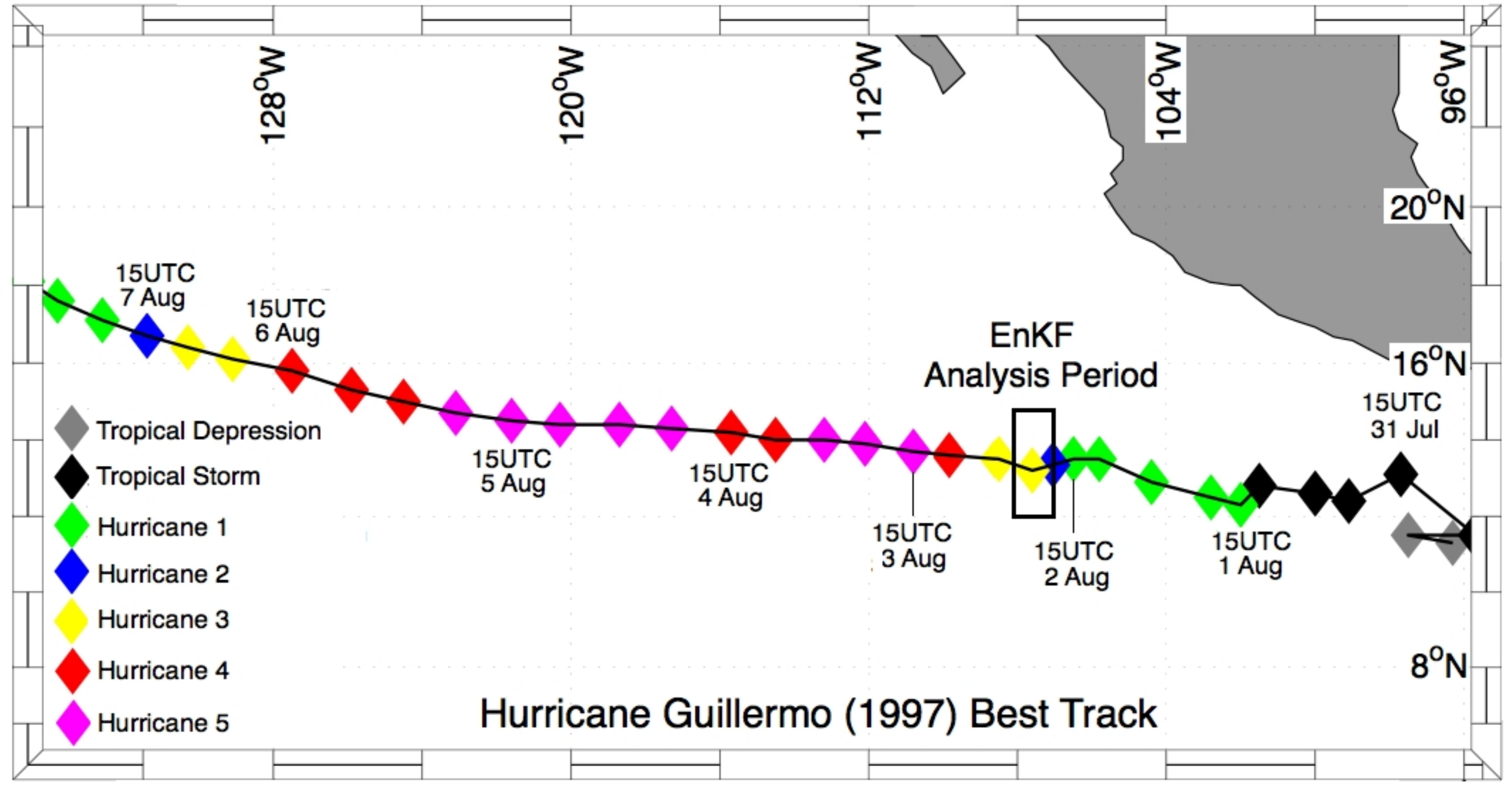}
   \caption{Best Track for Hurricane Guillermo (1997). Hurricane
   Intensity is color-coded based on the Saffir-Simpson scale with
   legend shown on the bottom left of the figure. Data courtesy of the
   Tropical Prediction Center (TPC), NOAA.  The EnKF analysis period
   is denoted by the small black rectangle.  }
   \label{fig:GuillermoTrack}
\end{figure}

One source of uncertainty within hurricane models come from
parameters, which dominate the long-term behavior of the model.
To explore whether these long-term uncertainties can be reduced, model
parameters associated with both environmental and physical forcings
are estimated for sheared hurricane Guillermo (1997, see
Fig.~\ref{fig:GuillermoTrack}) through the use of the ensemble Kalman
filter (EnKF).  Specifically, four parameters describing momentum
sinks and moisture sources in the planetary boundary-layer, the
unresolved transport of these quantities away from the boundary-layer,
and a parameter associated with describing the wind shear impacting
Guillermo will be estimated.  One of the key points of this paper is
to illustrate that parameter uncertainty contributes significantly to
the overall long-term uncertainty in a hurricane simulation.

Various authors (\cite{Anderson2001}; \cite{AnnanEtAl2005};
\cite{HackerSnyder2005}; \cite{AksoyEtAl2006a}; \cite{AksoyEtAl2006b};
\cite{TongXue2008}; \cite{HuEtAl2010}; \cite{NielsenEtAl2010}) have
documented the ability of the EnKF procedure to simultaneously
evaluate model state and parameters. In these papers, the parameters
are included as part of the model state in the assimilation.  This
combination of evolving elements (model variables) and non-evolving
elements (model parameters) within the
analysis introduces some difficulties for parameter estimation, such
as parameter collapse and assimilation divergence. To mitigate these
difficulties, the parameters are inflated to a prespecified variance,
so as to avoid the collapse and keep a reasonable spread in the
parameters. These techniques have proven to be effective for parameter
estimation, but they require adjustments and tuning of the inflation
to obtain good estimates.
The approach in our current paper differs from previous work in the
sense that the parameters and model state are not combined in the
assimilation procedure in order to estimate the parameters.  In this
work we use the EnKF as a tool to only estimate key model parameters
for a given time-distributed observational data set. Furthermore,
since the parameters are assumed to be non-evolving, they are
estimated independently from each observational data set in time. Once
the parameters are estimated at each time instance where observations
are available, a final estimate is obtained by computing the time
average value. A key aspect is to explore if estimating parameters
through EnKF data assimilation can improve model simulation,
especially for such a highly non-linear problem as a hurricane. To
test the applicability and viability of this approach, a
twin-experiment is performed for a hurricane model, where results
indicate that the correct parameter values are recovered when
considering sufficient observational data.  It must be noted that the
parameter estimates presented in the current work depends on two
important factors: the numerical model being utilized, and the data
set being assimilated. Nevertheless, this technique is applicable for
estimating model parameters for any model and data set, as long as the
parameters have a strong connection to the type of observational data
being used to estimate them.

One of the best dual-Doppler radar data sets (\cite{ReasorEtAl2009};
\cite{SitkowskiBarnes2009}) obtained within a hurricane will be used
to estimate the parameters through assimilation.  Some unique aspects
of Guillermo's dual-Doppler radar data were that data from 10 flight
legs over a six hour time period were individually processed to
address temporal variability and both derived fields of horizontal
winds and latent heating (\cite{GuimondEtAl2011}) were constructed from
the flight legs. Taken together this observational data will be used
to quantify temporal variability in the four parameter estimates along
with how these estimates change depending on which observational field
is utilized.  Note that assessing the temporal variability of the four
parameters is important with regard to addressing the effectiveness of
parameter estimation within this highly nonlinear system.

An important test of the viability of the particular approach and data
used to estimate the parameters, is their ability to improve the
solution of a hurricane model. To investigate this issue, three
simulations were run using temporally averaged values of the
parameters obtained from either derived observational field,
(horizontal winds or latent heat) or both fields. Hence, the final
point of this paper will be to illustrate whether or not the parameter
estimates improved overall model predictability with regard to the
type of observations being assimilated.

The paper is organized as follows;
Section~\ref{sec:ParameterEstimationModel} first describes the
predictive component of the parameter estimation model including the
analytic equation set of the hurricane model, the four model
parameters present within this predictive model, the discretization,
and a brief overview of the EnKF data assimilation method.  In
Section~\ref{sec:EnsembleSimulation} we describes various aspects of
the parameter estimation model setup including the ensemble setup and
the processing of the derived observational data fields.
Section~\ref{sec:Results} is broken up into three sub-sections with
each section showing results relating to the three major points of
this paper. A summary and final remarks are presented in
Section~\ref{sec:SummaryConclusions}.

\section{Parameter estimation model}
\label{sec:ParameterEstimationModel}
The parameter estimation model is comprised of a predictive model and
an parameter estimation model. The chosen predictive model is the
Navier-Stokes equation set coupled to a bulk cloud model with the four
parameters utilized within this model, whereas the parameter
estimation model employs the EnKF data assimilation.  In addition to
describing their representative analytical representations,
discussions regarding their respective discrete formulations will also
be presented in this section.  

\subsection{Predictive model}
\label{sec:PredictiveModel}

\subsubsection{Navier-Stokes equation set}
\label{sec:Navie-Stokes}

Since the analytical equation set representing the momentum, energy,
and mass of the gas phase is identical to that utilized in  
\cite{ReisnerJeffery2009}, RJ hereafter, interested
readers can examine this manuscript for details regarding its formulation.
Likewise the primary difference between the equation set
found in that paper and the current equation set are additional
terms associated with buoyancy in the vertical
equation of motion related to various
hydrometers, i.e., 
$g({\rho}'+\rho_c+\rho_r+\rho_i+\rho_s+\rho_g)$ where
$g$ is gravity, ${\rho}'$ is a density perturbation,
$\rho_c$ is the cloud water density, 
$\rho_r$ is the rain water density, 
$\rho_i$ is the ice water density, $\rho_s$ is
the density of snow, and $\rho_g$ is the graupel density, with
the next section briefly describing the bulk microphysical model used
to predict the evolution of the hydrometers. 

\subsubsection{Bulk microphysical model}
\label{sec:MicrophysicalModel}
The mass conservation equation for a given particle type,
${\rho}_{part}=\rho_c,\rho_r,\rho_i,\rho_s,\rho_g$, within the bulk
microphysical model can be written as follows
\begin{equation}
   {\partial (\rho_{part}) \over \partial t} +
   {\partial [(u^{i'}-wfall_{part} \delta_{i'3}) \rho_{part}] \over \partial x^{i'}}
    = fdensity_{part}
   +{\partial F^{i'}_{\rho_{part}} \over \partial x^{i'}},
   \label{micro.rho}
\end{equation}
where $u^{i'}$ are fluid velocities in each spatial direction
and the last term represents the turbulent diffusion of
a particle type using a diffusion coefficient diagnosed from a 
turbulence kinetic energy (TKE) equation, 
see Eq. 3 in \cite{ReisnerJeffery2009}.

The conservation equation for either cloud droplet number
($N_c$) or ice particle number ($N_i$), $N_{part}=N_c,N_i$, can be
written as follows
\begin{equation}
   {\partial (N_{part}) \over \partial t} +
   {\partial [(u^{i'}-wfall_{part} \delta_{i'3}) N_{part}] \over \partial x^{i'}}
    =fnumber_{part}+{\partial F^{i'}_{N_{part}} \over \partial x^{i'}},
   \label{micro.nt}
\end{equation}
where $wfall_{part}$, $fdensity_{part}$, and $fnumber_{part}$
represent the fall speed, density, and number sources or sinks from
the bulk microphysical model for a given particle type, a hybrid of
the activation and condensation model found in
\cite{ReisnerJeffery2009} together with
all of the other relevant bulk parameterizations found in
\cite{thompson}. Note, because of
significant differences in the particle distributions between winter
storms and hurricanes, the slope-intercept formulas were modified
following \cite{mcfarq}.

\subsubsection{Parameters of interest}
\label{sec:Parameters}
The initialization of the environmental or background horizontal
homogeneous potential temperature, water vapor, and total gas density
fields for all Guillermo simulations was achieved by examining
vertical profiles from ECMWF analyses obtained near the time period of
the dual-Doppler radar data (1830 UTC 2 August to 0030 UTC 3 August
1997) and using a representative composite. Though some uncertainty
exists within the thermodynamic fields with regard to the actual
environment versus the perturbed environment obtained from the ECMWF
soundings, the impact of this uncertainty was deemed to be smaller
than that associated with the momentum fields, i.e., the simulated
vortex is sensitive to small changes in wind shear.
%
%
So to quantify this sensitivity, the horizontal velocity fields,
$u^{1'}$ and $u^{2'}$, were initialized as follows
\begin{eqnarray}
   u^{1'}(x^{3'}) &=& \phi_{shear} [ecmwf_u(x^{3'})+1.5],
   \label{eq:u1}\\
   u^{2'}(x^{3'}) &=& \phi_{shear} [ecmwf_v(x^{3'})-1.5],
   \label{eq:u2}
\end{eqnarray}
where $\phi_{shear}$ is a tuning coefficient that determines the shear
impacting hurricane Guillermo within a range of 0 and 1, 
$x^{3'}$ is the height, and $ecmwf_u$
and $ecmwf_v$ represent mean soundings calculated from the ECMWF
analyses.

Given the delicate balance in nature that is needed for a sheared
hurricane to intensify, it is not entirely obvious whether numerical
models, that are necessarily limited in resolution, can accurately
represent boundary processes that are responsible for supplying water
vapor to eyewall convection. The accurate representation of
boundary-layer processes implies the model has been somewhat “tuned”
to represent the impacts of waves, sea spray, and air bubbles within
the water; likewise the accurate treatment of energy release in
eyewall convection implies that the upward movement of, for example,
moisture is being reasonably simulated by the hurricane model.

%
To examine this uncertainty the diffusion coefficient for surface
momentum calculations was specified as follows
\begin{equation}
\kappa=\kappa_{surface friction} 
\tanh \left({{\bf V_h} \over 80}\right),
\end{equation}
where $\kappa_{surface friction}$ is a tuning coefficient that ranges
from 0.1 to 10 m$^2$ s$^{-1}$ and ${\bf V_h}$ is the near surface
horizontal wind speed. A no-slip boundary condition was utilized in
the horizontal momentum equations ($u^{1'}$ =$u^{2'}$=0) with the
magnitude of $\kappa_{surface friction}$ the determining factor with
regard to the impact of this boundary condition on the intensity and
structure of Guillermo. Note, unlike for the horizontal momentum
equations, all scalar equations use a diffusion coefficient estimated
from the $TKE$ equation within calculations of surface fluxes.

Another uncertain boundary-layer process that has a significant impact
on intensification rate is surface moisture availability and the
unresolved vertical transport of this water vapor with the first term,
$q^s_v$, being formulated as follows
\begin{equation} 
{q^s_v}=q_{vs} \left( 0.75+{q_v}_{surface} \right) \tanh \left({{\bf V_h} \over 30}\right),
\end{equation}
where $q_{vs}$ is the saturated vapor pressure over water and
${q_v}_{surface}$ is a tuning coefficient that ranges in value from
0.0 to 0.2. This term enters into surface diffusional flux
calculations of water vapor, $F^{3'}_{q_v}$, in discrete form as follows
\begin{equation}
F^{3'}_{q_v}=\kappa { {q^1_v - q^s_v} \over {0.5 \Delta x^{3'}}},
\end{equation}
where $q^1_v$ is the specific humidity of the first grid cell in the
vertical direction.  To
address the uncertainty associated with the turbulent transport of water
vapor (and all other fields) from the surface to the free atmosphere
the turbulent length scale was modified as follows 
\begin{equation}
L^m_s={\phi}_{turb} L_s,
\end{equation}
where the tuning coefficient, ${\phi}_{turb}$, ranged from 0.1 to 10, and
the eddy diffusivity now being $\kappa=0.09 L^m_s \sqrt {TKE} $.

\subsubsection{Discrete model}
\label{sec:DiscreteModel}
The discrete model for the Navier-Stokes equation set and the bulk
microphysical model closely follows what was described in section 2c
of RJ.  This discrete equation set formulated on an A-grid can utilize
a variety of time-stepping procedures with the current simulations
using a semi-implicit procedure
(\cite{ReisnerEtAl2005}).  The
advection scheme used to advect gas and various cloud quantities was
the quadratic upstream interpolation for convective kinematics
advection scheme including estimated streaming terms
(QUICKEST,~\cite{leonard}) with these
quantities having the possibility of being limited by a flux-corrected
transport procedure (\cite{zalesak1}).

%
The domain spans 1200 km in either horizontal direction and 21 km in
the vertical direction. The stretched horizontal mesh employing 300
grid points has the highest resolution of 1 km at the center of the
mesh and lowest resolution of 7 km at the model edges. Because of the
addition of a mean wind intended to keep the vortex centered in the
middle of the domain, the coarsest resolution resolving the highest
wind field of Guillermo is approximately 2 km.  The stretched vertical
mesh is resolved by 86 grid points with highest resolution of 50 m at
the surface and lowest of 500 m at the model top.

\subsection{Parameter estimation Model}
\label{sec:ParameterEstimation}
In this section the ensemble Kalman filter (EnKF) method is briefly
described. In the current work, the EnKF is mainly utilized to
estimate the model parameters.

\subsubsection{Parameter estimation with EnKF}
\label{sec:ParameterEstimationEnKF}
The ensemble Kalman Filter is a Monte Carlo approach of the Kalman
filter which estimates the covariances between observed variables and
the model state variables through an ensemble of predictive model
forecasts. The EnKF was first introduced by \cite{Evensen1994} and is
discussed in detail in \cite{EvensenVanLeeuwen1996} and in
\cite{HoutekamerMitchell1998}.
For the current study, only the parameters will be estimated, not the
state vector of the model. The EnKF procedure is directly applied to
the parameters, i.e., the state vector contains only the parameter
values. Nevertheless, the model covariance matrix is still required
for the innovation with observations. The following EnKF description
will be concern with model parameters.

Let $\mathbf{p} \in \mathbb{R}^{\ell}$ be a vector holding the
different model parameters, and $\mathbf{x}^{f} \in \mathbb{R}^{n}$ be
the model state forecast. Let $\left( \mathbf{p}_{i},
\mathbf{x}^{f}_{i} \right)$ for $i=1 \ldots N$ be an ensemble of model
parameters and state forecasts, and $\mathbf{y}^{o} \in
\mathbb{R}^{m}$ a vector of $m$ observations, then the estimated
parameter values $\mathbf{p}^{a}_{i}$ given by the EnKF equations are
\begin{eqnarray}
   \mathbf{p}^{a}_{i} &=& \mathbf{p}_{i} + \tilde{\mathbf{K}}\left(
   \mathbf{y}^{o}_{i} - \mathbf{H} \mathbf{x}^{f}_{i} \right), \qquad
   i=1,\ldots,N \label{eq:EnKF1}\\
   \tilde{\mathbf{K}} &=& \mathbf{C}^{T}\mathbf{H}^{T} \left(
   \mathbf{H}\mathbf{P}^{f}\mathbf{H}^{T} + \mathbf{R} \right)^{-1},
   \label{eq:EnKF2}
\end{eqnarray}
where the matrix $\tilde{\mathbf{K}} \in \mathbb{R}^{\ell \times m}$
is a modified Kalman gain matrix (see Appendix A), $\mathbf{P}^{f} \in
\mathbb{R}^{n \times n}$ is the model forecast covariance matrix,
$\mathbf{C} \in \mathbb{R}^{n \times \ell}$ is the
cross-correlation matrix between the model forecast and parameters,
$\mathbf{R} \in \mathbb{R}^{m \times m}$ is the observations
covariance matrix, and $\mathbf{H} \in \mathbb{R}^{m \times n}$ is an
observation operator matrix that maps state variables onto
observations. In the EnKF, the vector $\mathbf{y}^{o}_{i}$ is a
perturbed observation vector defined as
\begin{equation}
   \mathbf{y}^{o}_{i} = \mathbf{y}^{o} + \varepsilon_{i},
   \label{eq:PerturbedObs}
\end{equation}
where $\varepsilon_{i} \in \mathbb{R}^{m}$ is a random vector sampled
from a normal distribution with zero mean and a specified standard
deviation $\sigma$. Usually $\sigma$ is taken as the variance or error
in the observations.

One of the main advantages of the EnKF is that the model forecast
covariance matrix is approximated using the ensemble of model
forecasts,
\begin{equation}
   \mathbf{P}^{f} \approx \frac{1}{N-1} \sum^{N}_{i=1} \left(
   \mathbf{x}^{f}_{i} - \bar{\mathbf{x}}^{f} \right)\left(
   \mathbf{x}^{f}_{i} - \bar{\mathbf{x}}^{f} \right)^{T},
   \label{eq:ModelCovariance}
\end{equation}
where $\bar{\mathbf{x}}^{f} \in \mathbb{R}^{n}$ is the model forecast
ensemble average. The use of an ensemble of model forecast to
approximate $\mathbf{P}^{f}$ enables the evolution of this matrix for
large non-linear models at a reasonable computational cost.
Additionally, the cross-correlation matrix $\mathbf{C}$ is defined as
\begin{equation}
   \mathbf{C} = 
   \frac{1}{N-1} \sum^{N}_{i=1} \left(
   \mathbf{x}^{f}_{i} - \bar{\mathbf{x}}^{f} \right)\left(
   \mathbf{p}_{i} - \bar{\mathbf{p}} \right)^{T},
   \label{eq:CrossCorrelation}
\end{equation}
where $\bar{\mathbf{p}} \in \mathbb{R}^{\ell}$ is the parameter
ensemble average.

For our particular implementation, the system of equations
\eqref{eq:EnKF1}-\eqref{eq:EnKF2} is rewritten as
\begin{eqnarray}
   && \left( \mathbf{H}\mathbf{P}^{f}\mathbf{H}^{T} + \mathbf{R}
   \right) \mathbf{z}_{i} = \left( \mathbf{y}^{o}_{i} -
   \mathbf{H}\mathbf{x}^{f}_{i} \right) \label{eq:EnKF3} \\
   && \mathbf{p}^{a}_{i} = \mathbf{p}_{i} +
   \mathbf{C}^{T}\mathbf{H}^{T}\mathbf{z}_{i}, \label{eq:EnKF4}
\end{eqnarray}
for $i=1,\ldots,N$, where $\mathbf{z}_{i} \in \mathbb{R}^{m}$ is the
solution of the linear system \eqref{eq:EnKF3} for ensemble $i$. For
our implementation, the observation covariance matrix $\mathbf{R}$ is
taken as a diagonal matrix, with $\sigma$ in its main diagonal.

\subsubsection{Considerations for parameter estimation with EnKF}
\label{sec:ConsiderationsEnKF}

Several studies have utilized the EnKF data assimilation to
simultaneously estimate the model state and parameter. Among them are
the studies by \cite{AksoyEtAl2006b}, \cite{AksoyEtAl2006a},
\cite{TongXue2008}, \cite{HuEtAl2010}, and \cite{NielsenEtAl2010}.
Typically, the parameters are included as part of the model state in
the assimilation. This evolving of dynamical and non-evolving
elements within the analysis introduces some difficulties for
parameter estimation, such as parameter collapse and assimilation
divergence. To mitigate these difficulties, the parameters are
inflated to a prespecified variance, so as to avoid the collapse and
keep a reasonable spread in the parameters. These techniques have
proven to be effective to estimate parameter, but they require
adjustments and tuning of the inflation to obtain good estimates.

The particular approach taken in this work is to use the EnKF as a
tool to only estimate the model parameters using the available data.
This approach is significantly different from the studies mentioned
above in the sense that only the parameters are estimated with the
EnKF, that is, the model state is not being estimated. The motivation
behind this approach is that model parameters are assumed to be
constant, they do not evolve through the model, although
they affect the dynamics of the solution. For this reason, determining
parameters can be viewed as a stationary or static optimization. Our
objective is to estimate a constant parameter value for the given data
set, over the given time window. To achieve this, the assimilation of
the data is performed on each time instance, where observations are
available, independently.  The reasoning behind this technique is to
treat the parameters as constants and non-evolving elements in the
model, hence for each time period we compute an estimated parameter
value.

The procedure used to estimate the parameters is the following: Let
$t_{1},\ldots,t_{k}$ be the time instances where observations are
available. For each time instance $t_{j}$,~$j=1,\ldots,k$, the EnKF
data assimilation (equations \eqref{eq:EnKF3}-\eqref{eq:EnKF4})
provides parameter estimates for the ensemble,
$\mathbf{p}^{a}_{i}\left( t_{j} \right)$, $i=1\ldots N$. A final
parameter estimate is then computed by first taking the ensemble
average and then the time average of the parameters, that is
\begin{equation}
   \mathbf{p}^{a} = \frac{1}{k}\sum^{k}_{j=1} \left[ \frac{1}{N}
   \sum^{N}_{i=1} \mathbf{p}^{a}_{i}\left( t_{j} \right) \right]   
   \label{eq:ParameterEstimate}
\end{equation}
One advantage is that this approach avoids the problem of parameter
collapse and filter divergence, since the data assimilation is used to
estimate the parameters at each time instance independently.
Additionally, since the state is not being updated in the
assimilation, and only the parameter are being estimated, localization
is not required for the EnKF.


The number of available data points for Hurricane Guillermo is about
$200,000$ at any given time. This rich data set can be used to
investigate a number of aspects of hurricane intensification. In order
to exploit this data set for assimilation, we used an efficient
matrix-free EnKF algorithm developed by \cite{GodinezMoulton2012}. The
algorithm works by efficiently solving the linear system
\eqref{eq:EnKF3} using a solver based on the Sherman-Morrison formulas.
In their paper, \cite{GodinezMoulton2012} show that this algorithm is
more efficient than traditional implementations of the EnKF, by
several orders of magnitude, and enables the assimilation of vast
amounts of data. Additionally, the algorithm provides an analysis that
is qualitatively and quantitatively the same as more traditional
implementations. The reader is referred to the work of
\cite{GodinezMoulton2012} for more details.

\section{Parameter estimation model setup}
\label{sec:EnsembleSimulation}
This section describes the three necessary steps to conduct the
parameter estimates: the processing of derived observational data
fields; the setting up of the ensemble; and the setup of the EnKF
system.

\subsection{Derived Doppler data fields}
\label{sec:data}

The primary driver of a hurricane is the release of latent heat in
clouds, which arises mainly from condensation.  Latent heat cannot be
observed directly and instruments, such as Doppler radars, only
measure the reflectivity and radial velocity of precipitation
particles averaged over the pulse volume.  As a result, retrievals of
dynamically relevant quantities (e.g. the Cartesian wind components
and latent heat) are required.  Guillermo's 3-D wind field was
retrieved using a variational approach on a system of equations that
includes the radar projection equations, the anelastic mass continuity
equation, and a Laplacian filter (\cite{GaoEtAl1999};
\cite{ReasorEtAl2009}).  This wind field and estimates of the
precipitation water content (derived from the reflectivity
measurements) are used to retrieve the latent heat of
condensation/evaporation following \cite{GuimondEtAl2011}.  There are two
main steps in the latent heat retrieval algorithm:  (1) determine the
saturation state at each grid point in the radar analysis using the
precipitation continuity equation; and (2) compute the magnitude of
heat released using the first law of thermodynamics and the vertical
velocity estimates described in \cite{ReasorEtAl2009}.  There are
several potential sources of error in the latent heat retrievals and a
detailed treatment of these errors can be found in
\cite{GuimondEtAl2011}.
Here, we summarize the most relevant information.

The uncertainty in the latent heat retrievals reduces to uncertainties
in two main fields:  reflectivity and vertical velocity.
\cite{GuimondEtAl2011} were able to reduce and/or document the uncertainty
in these fields to a level where the latent heat retrievals have a
reasonably acceptable accuracy.  For example, \cite{GuimondEtAl2011} focus
on the inner portion of the eyewall and often use two aircraft to
construct the radar analyses (\cite{ReasorEtAl2009}), which reduce the
effects of attenuation.  In an attempt to correct the known
calibration bias in the NOAA P-3 Tail radar reflectivity, 7 dB was
added to the fields (John Gamache and Paul Reasor, personal
communication).  More importantly, however, the reflectivity is only
used to determine the condition of saturation in the latent heat
retrieval.  Thus, the algorithm is not dependent on the precise value
of the reflectivity rendering the retrievals somewhat insensitive to
errors (\cite{GuimondEtAl2011}).

\begin{figure}
   \centering
   \includegraphics[width=\linewidth]{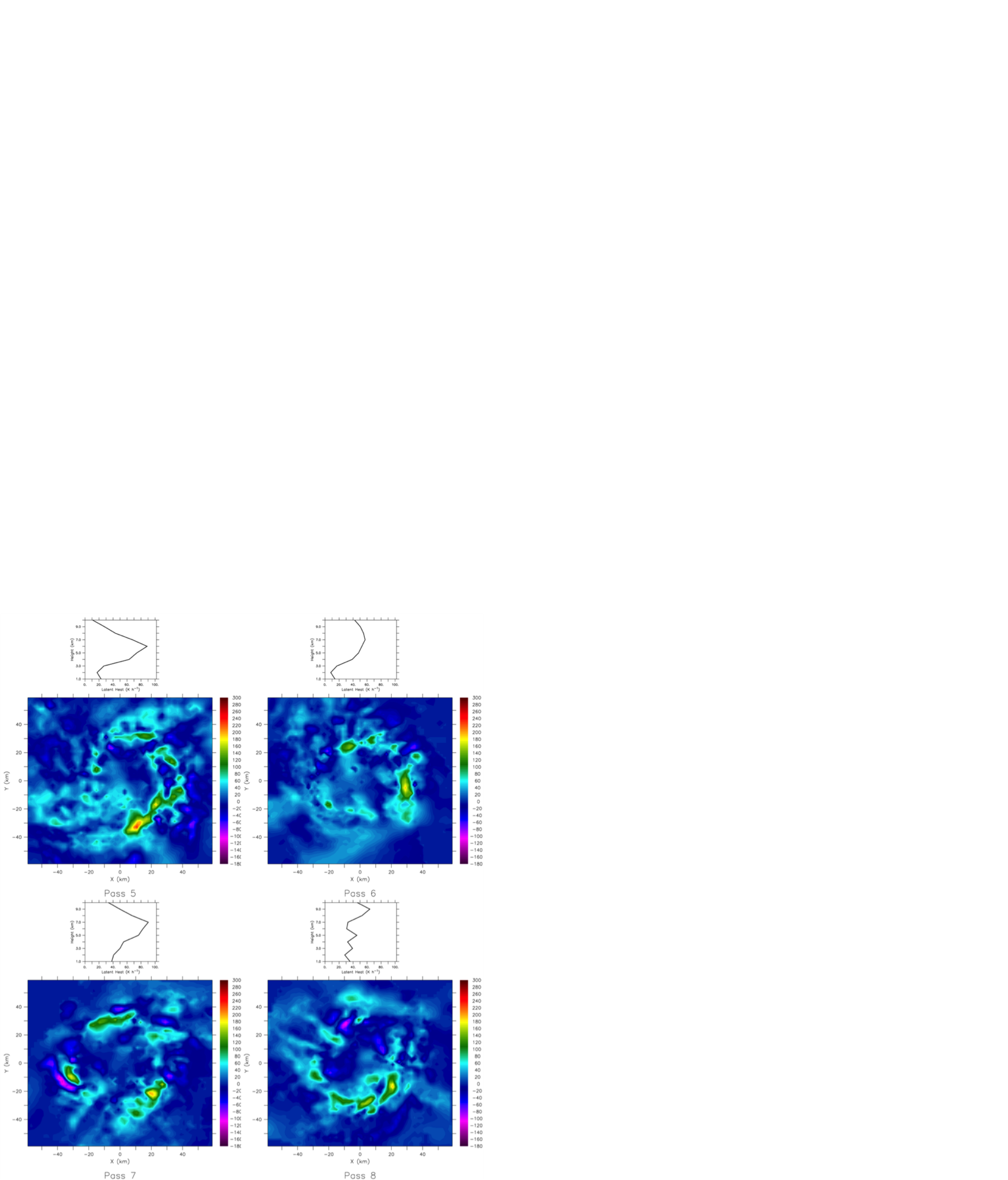}
   \caption{Horizontal views (averaged over all heights) of the latent
   heating rate (K~h~$^{-1}$) of condensation/evaporation retrieved from
   airborne Doppler radar observations in Hurricane Guillermo (1997)
   at four select times of the dual-Doppler radar data, i.e., pass 5
   corresponds to 2117 UTC from Fig. 6 of \cite{ReasorEtAl2009}.
   Note that grid points without latent
   heating were assigned zero values after the vertical averaging.
   The vertical profile of the azimuthal mean latent heating rate at
   the RMW (30 km) is shown above each contour plot.  The first level
   of data is at 1 km due to ocean surface contamination. } 
   \label{fig:3DLatentHeat}
\end{figure}
The uncertainties in the magnitude of the retrieved heating are
dominated by errors in the vertical velocity.  Using a combination of
error propagation and Monte Carlo uncertainty techniques, biases are
found to be small, and randomly distributed errors in the heating
magnitudes are 16\% for updrafts greater than 5~m~s$^{-1}$ and 156\%
for updrafts of 1~m~s~$^{-1}$ (\cite{GuimondEtAl2011}).  Even though
errors in the vertical velocity can lead to large uncertainties in the
latent heating field for small updrafts/downdrafts, in an integrated
sense, the errors are not as drastic. Figure~\ref{fig:3DLatentHeat}
(from \cite{GuimondEtAl2011}) shows example horizontal views (averaged
over all heights) of the latent heating rate of
condensation/evaporation for four of the ten aircraft sampling periods
of Guillermo.

For the assimilation, only latent heat data where there is a non-zero reflectivity value are incorporated into the data set. The errors
for the retrieved wind fields are set to $5.0$ to $6.0$\%, and the
errors in the retrieved latent heating, as a percentage, are specified
following \cite{GuimondEtAl2011}
\begin{equation}
   \delta \mathbf{y}^{o}_{lh} = \left| \frac{\delta
   \mathbf{w}}{\mathbf{w}}\right| \times 100
   \label{eq:LatentHeatError}
\end{equation}
where $\delta \mathbf{w} = 1.56$~m~s~$^{-1}$ represents the overall
uncertainty in the vertical wind velocity field $\mathbf{w}$
\cite{ReasorEtAl2009}. It is
worth to notice that these errors are sometimes overestimated or
underestimated. Thus in an integral sense, the errors are not so
drastic, the bias is only of $+0.16$~m~s~$^{-1}$.

\subsection{Guillermo ensemble setup}
\label{sec:ModelSetupStructure}

Since the primary goal is to examine the impact of various model
parameters containing high uncertainty on the intensity and structure
of Guillermo, but not the track, all simulations comprising the
ensemble have been undertaken in which a mean wind of 1.5 m s$^{-1}$
was added or subtracted to  the respective environmental wind
components to prevent the movement of Guillermo from a region
containing high spatial resolution found in the domain center.
Specifically this high resolution patch in Cartesian space, $\Delta
x^{1'}_c$ and $\Delta x^{2'}_c$, is defined as follows
\begin{eqnarray}
   \Delta x^{1'}_c &=& 6000 {\sin}^2 ({\phi}_g x^{*})+1000, \label{eq:deltax1}\\
   \Delta x^{2'}_c &=& 6000 {\sin}^2 ({\phi}_g y^{*})+1000, \label{eq:deltax2}
\end{eqnarray}
where ${\phi}_g={\pi \over {Ngp_{i'}}}$ determines how quickly the
grid spacing changes from 7 km near the model edges to 1 km near the
center with $Ngp_{i'}$ the number of grid points in either direction
and  $x^{*}$,$~y^{*}$ represent grid values for a normalized grid with
a domain employing $0.5Ngp_{i'}$ grid points away from a center
location in which $x^{*}=y^{*}=0$.  Like the horizontal direction, the
vertical direction also employs stretching with highest resolution
near the ocean boundary, approximately 50 m, and coarsest near the
model top, 500 m, with 86 vertical grid points being utilized to
resolve a domain extending upwards to 21 km. Note, because of the
relatively high vertical spatial resolution, time step size was
limited to 1 s to avoid any instabilities associated with exceeding
the advective Courant number limit. 

The ensemble is generated by perturbing only the four parameters
discussed in Section \ref{sec:Parameters}, which are ${\phi}_{shear}$,
${\kappa}_{surfacefriction}$, ${q_v}_{surface}$, and ${\phi}_{turb}$.
The parameter values are generated by the Latin hypercube sampling
technique with a uniform distribution, where each parameter is sampled
over a specified interval.  All ensembles have the same initial, where
the background fields are initialized as described in Section
\ref{sec:Parameters}, where the background winds that are independent
of the wind field associated with Guillermo are initialized using
ECMWF data. Whereas to initialize the wind field associated with
Guillermo a composite of the radar winds and a bogus vortex is
employed within a nudging procedure over a one hour time period.
%
%
%
Afterwards, the hurricane model is simulated in free mode for five
hours to become balanced with the parameters, and develop a hurricane
vortex. It is important to mention that at this stage the behavior of
the ensemble simulations are mainly dominated not by the initial
conditions, but by the parameter values. 
%
%
Hence utilizing derived fields from the same hurricane for a vortex
initialization does not bias the results of the parameter estimation
experiments, as long as the subsequent spin-up is sufficiently long
to allow the parameters dominate the long-term behavior of the
simulation.


\subsection{EnKF and Observation Selection}
\label{sec:EnKFSetup}
The EnKF data assimilation is used to estimate only the parameters of
interest.  The time distribution of the parameters is obtained by
assimilate each time period independently, as stated in Section
\ref{sec:ConsiderationsEnKF}. A final parameter value is obtained by
averaging in time the results of the assimilation. The particular
algorithm used for the parameter EnKF is a matrix-free algorithm
presented in \cite{GodinezMoulton2012}.

\begin{table}[t]
   \centering
   \begin{tabular}{rr}
      \hline\hline
      parameter & interval \\
      \hline
      surface moisture & $\left[0.05, 0.2\right]$ \\
      wind shear &  $\left[0.1, 1.0\right]$ \\
      turbulent length scale &  $\left[0.1, 10.0\right]$ \\
      surface friction &  $\left[0.1, 10.0\right]$ \\
      \hline
   \end{tabular}
   \caption{Initial parameter intervals for sampling with Latin
   Hypercube strategy using a uniform distribution.}
   \label{tab:ParameterIntervals}
\end{table}
\begin{figure}
    \centering
    \includegraphics[width=0.9\linewidth]{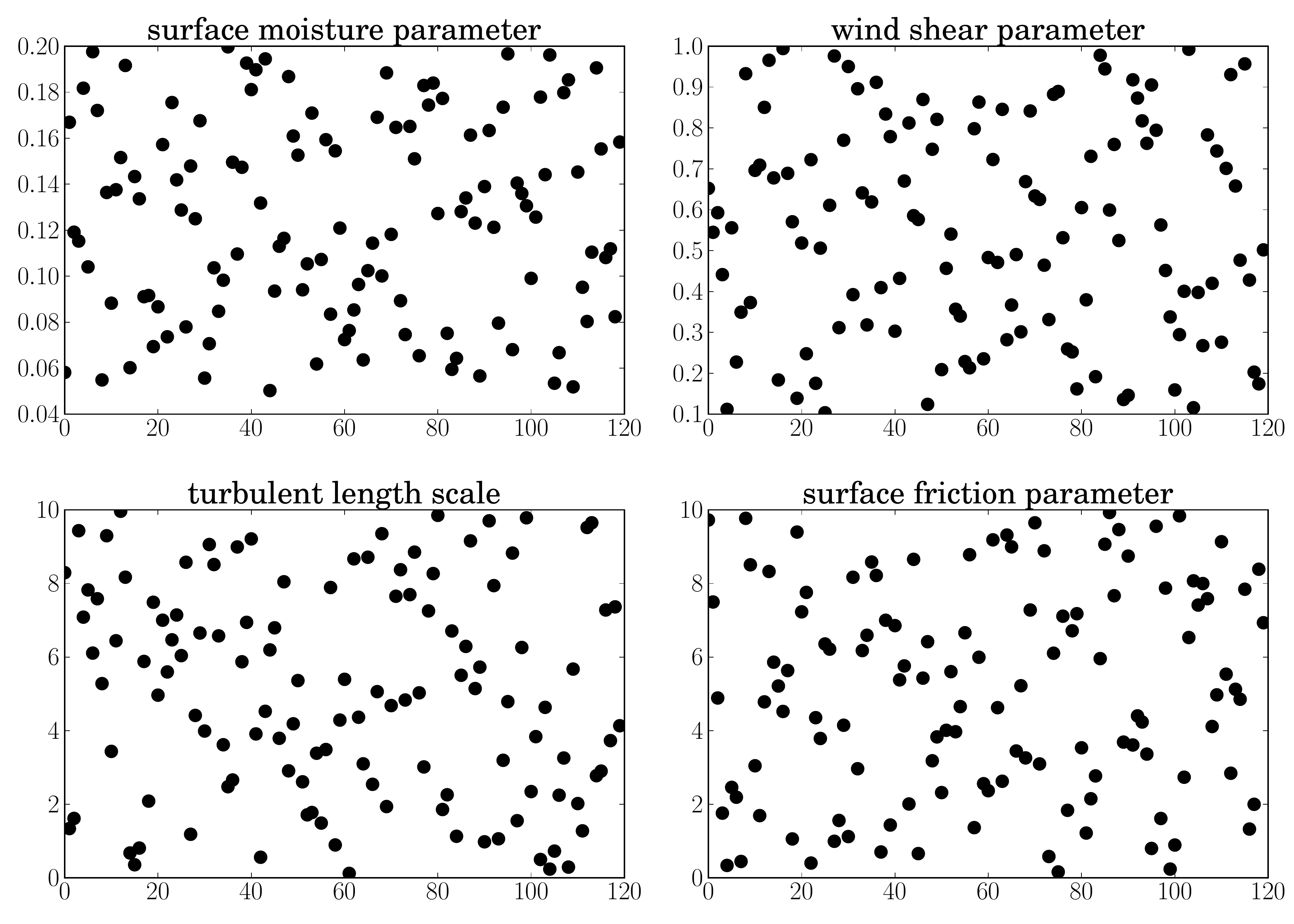}
    \caption{The parameter spread of the 120 ensemble members obtained by utilization of
    the Latin Hypercube sampling strategy within the limits shown in Table
    \ref{tab:ParameterIntervals}}.
    \label{fig:InitialParameterSpread}
\end{figure}
The initial set of parameter values are generated using a Latin
Hypercube sampling strategy with a uniform distribution.  The interval
for each parameter, shown in Table \ref{tab:ParameterIntervals} is
chosen according to the values mentioned in Section
\ref{sec:PredictiveModel}, which are based on prior sensitivity
simulations and from a physical intuition of their interaction with
the model.  Figure \ref{fig:InitialParameterSpread} shows the initial
distribution for each parameter of interest.

Although the optimal ensemble size for estimating reliable model
uncertainty is still under active research, an ensemble of 120 members
was deemed appropriate to capture essential model statistics for
parameter estimation.  Upon completion of the 120 member ensemble and
due to the small movement of simulated hurricanes within the ensemble,
information from each ensemble member was independently interpolated
to the center location of the grid such to be coincident with the
observations.  Note, only a portion of the computational domain was
utilized within the EnKF corresponding to the portion of the domain
that contains the derived fields of winds and latent heat. Finally,
after this interpolation step, all interpolated model and derived data
fields were read into the EnKF to conduct the parameter estimation for
a given time period.

The data selection procedure followed in this paper is similar to the
rank parameter-observation correlation procedure presented by
\cite{TongXue2008}. The parameter cross-correlation matrix,
$\mathbf{\mathbf{C}}$ in equation \eqref{eq:CrossCorrelation},
provides information of the correlation between the model state variables
and the parameters. By applying the observation
operator $\mathbf{H}$ to the model state, we have
\begin{equation}
   \tilde{\mathbf{C}} = \frac{1}{N-1} \sum^{N}_{i=1} \left( \mathbf{H}
   \mathbf{x}^{f}_{i} - \mathbf{H}\bar{\mathbf{x}}^{f} \right)\left(
   \mathbf{p}_{i} - \bar{\mathbf{p}} \right)^{T},
   \label{eq:CrossCorrelation2}
\end{equation}
which is a cross-correlation matrix 
between parameters and model state variables in observation space. The
correlations are then sorted and the observational data points that
are located in regions of largest correlations are selected for
assimilation. It is important to note that the regions of high
correlation can be identified from different model variables fields,
such as latent heat, horizontal winds, or reflectivity. This enables
the use one of these fields as a proxy for observation localization in
order to better capture the physical processes of interest.
\section{Results}
\label{sec:Results}
This section will highlight the three main points of this paper: 1)
the highly nonlinear interactions between the various parameters
leading to large variations in the simulated intensity and structure
of Guillermo among the 120 ensemble members; 2) the large amounts of
derived data, horizontal winds or latent heat, required to reasonably
estimate the four model parameters; and 3) the newly estimated
parameters lead to an overall reduction in the model forecast error.

\subsection{Ensemble spread and structure}
\label{sec:EnsembleStructure}
\begin{figure}
   \centering
   \includegraphics[width=\linewidth]{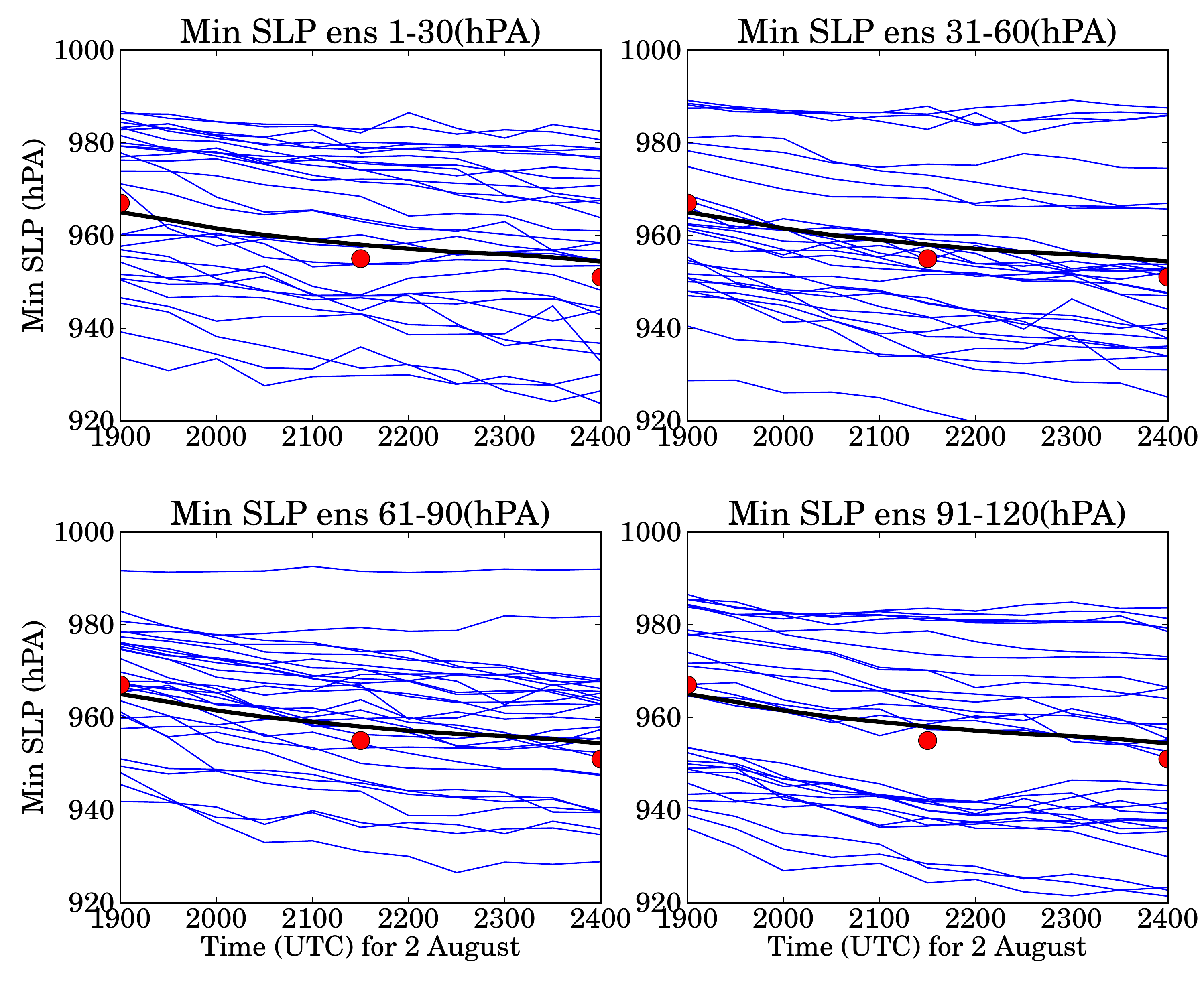}
   \caption{Minimum sea level pressure versus simulation time for each
   ensemble member (blue line), ensemble average (black line), and
   observations (red dots), for ensemble 1-30 (top left), ensemble
   31-60 (top right), ensemble 61-90 (bottom left), and ensemble
   91-120 (bottom right).}
   \label{fig:SLP2}
\end{figure}
Since the EnKF method depends on proper model statistics to optimally
determine the parameter values (\cite{Anderson2001}) this section will
illustrate this necessary variability across the various members of
the ensemble.  Likewise, for reasonable parameter estimation it is
important that the ensemble produces statistics that are within the
range of the observations and this is another aspect of the ensemble
that will be presented.  For example, Fig.~\ref{fig:SLP2} shows the
simulated pressure traces for the 120 ensemble members (blue lines),
the ensemble average (black line), and the 3-hourly observations from
the NHC advisories (red dots) with a large spread in hurricane
intensity being denoted within the ensemble.  Further, even though
this result may be somewhat fortuitous, the ensemble-averaged pressure
trace is in remarkably good agreement with observations with a
difference less than 5~hPa.  To highlight differences between the
ensemble average and a given member, in addition to displaying
ensemble average statistics, results from a select ensemble member,
i.e., member~44, that also reasonably reproduced the observed pressure
trace will be shown.

\begin{figure}
    \centering
    \includegraphics[width=\linewidth]{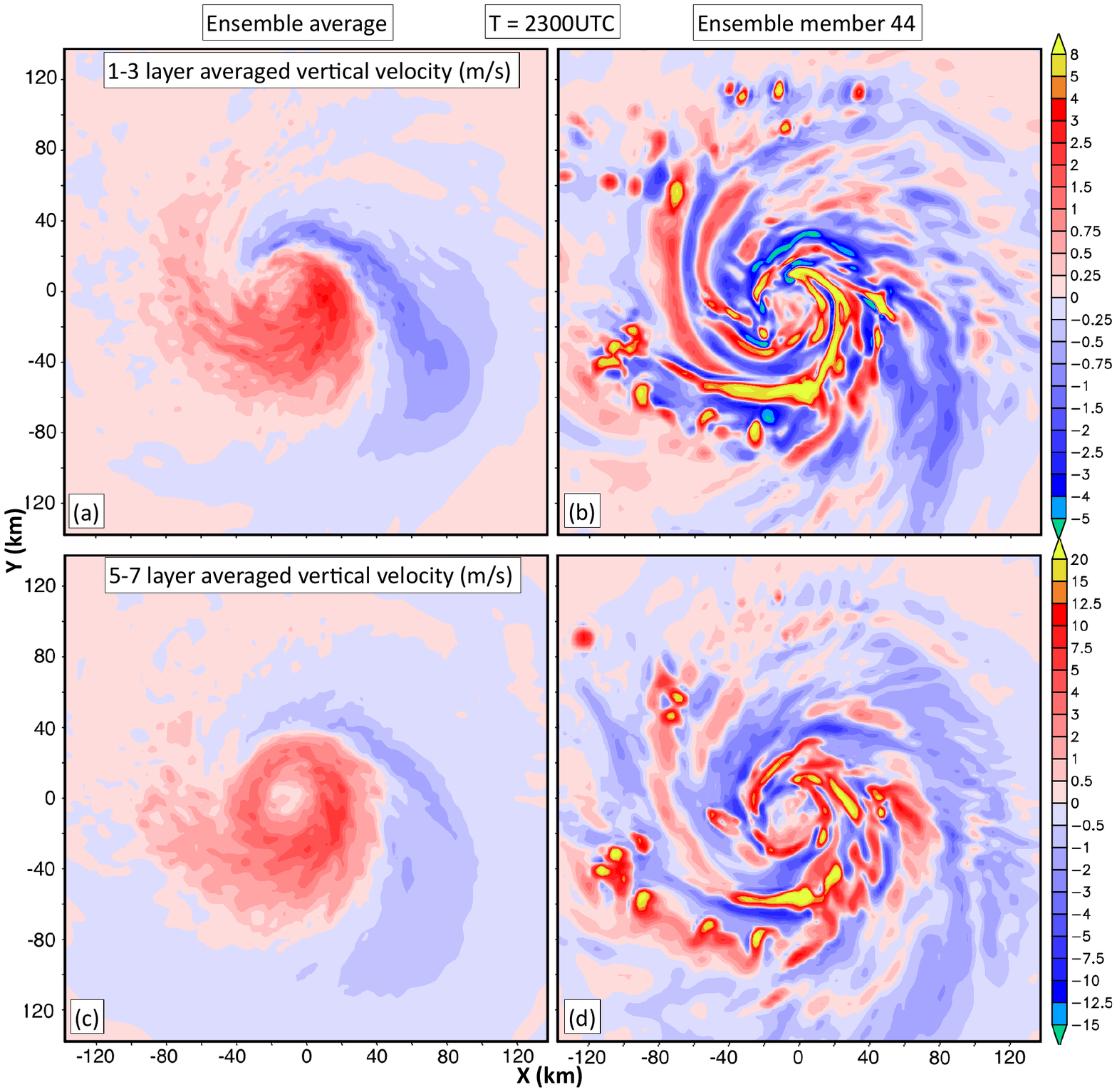}
    \caption{Ensemble average vertical motion fields 
     at 2300 UTC (11 hours into the simulations)
      averaged between 1-3 km (a) or 5-7 km (c). Corresponding
     layer-averaged vertical motions fields from ensemble member 44 
     between 1-3 km (b) or 5-7 km (d). }
    \label{fig:EnsembleAverageW39600}
\end{figure}
\begin{figure}
    \centering
    \includegraphics[width=\linewidth]{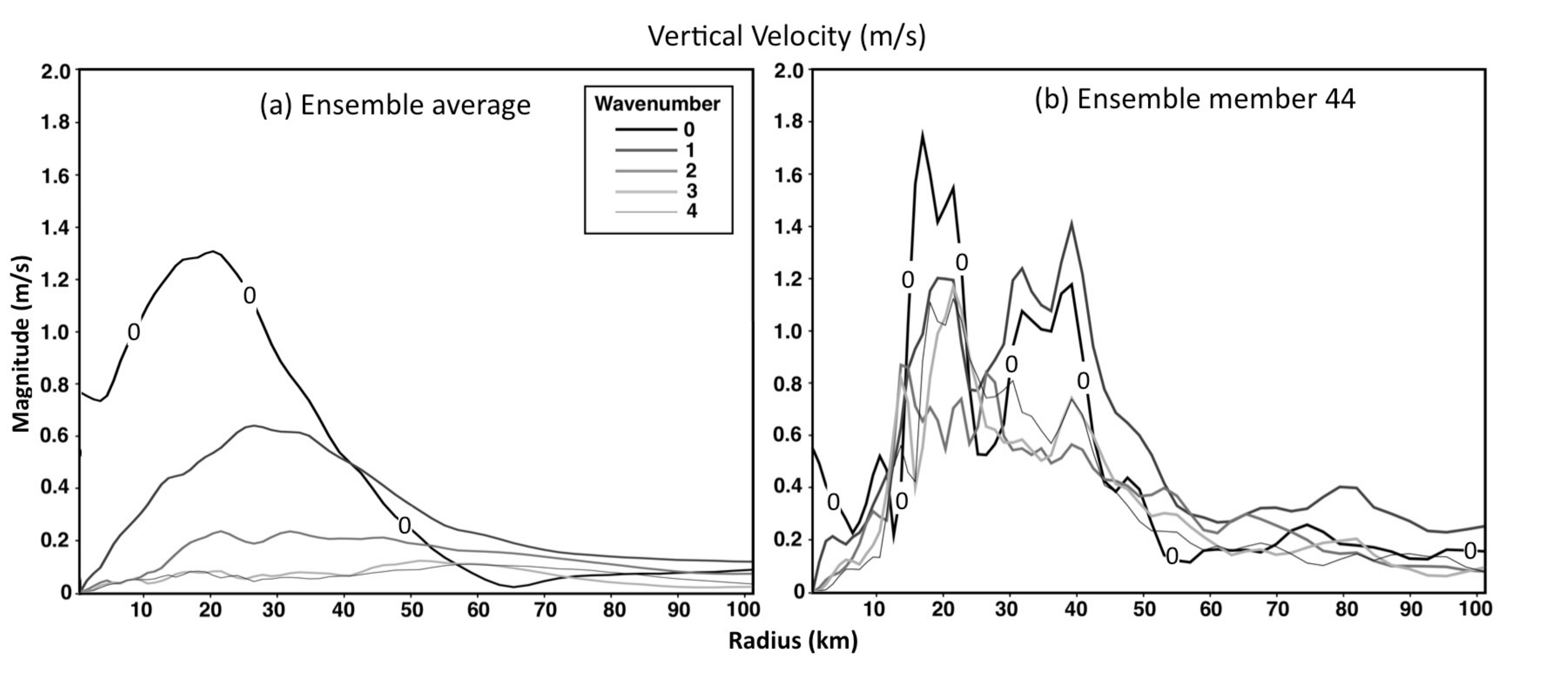}
    \caption{Time-averaged axisymmetric (black solid line) and
    azimuthal wavenumber-1–4 amplitudes of vertical velocity in the
    1–3-km layer within a 200 km radial distance from the storm
    center from the ensemble average (left figure) or ensemble
    44 (right figure). The averaging times are between 2230 UTC and 2300 UTC.
    }
    \label{fig:ForierW}
\end{figure}
Because hurricane Guillermo was embedded in an environment
characterized with vertical wind speed shear, its observed eyewall
horizontal structure was asymmetric with a dominant wavenumber 1 mode,
e.g., \cite{ReasorEtAl2009} and \cite{SitkowskiBarnes2009}.  To
illustrate the models ability to reproduce this asymmetry
Fig.~\ref{fig:EnsembleAverageW39600} shows both the ensemble-averaged
layer-averaged vertical velocities and the corresponding
layer-averaged fields from member~44 at two different layers. As
evident in this figure, the vortex in both the average sense and for
member~44 is asymmetric with a dominant wavenumber~1 mode being
readily apparent. Likewise, the impact of averaging across all members
is clearly evident in Fig.~\ref{fig:EnsembleAverageW39600} with a
significant smoothing and reduction in vertical motions being noted
with regard to the vertical motion fields produced by ensemble
member~44.  Furthermore, using a similar Fourier spectral
decomposition procedure as \cite{ReasorEtAl2009}
Fig.~\ref{fig:ForierW} reveals significant amounts of the vertical
motion fields being in wavenumber~0 and 1 components with the
magnitude of the wavenumber~1 vertical motion field from ensemble
member~44 reasonably agreeing with the observations (see Fig.~15a of
\cite{ReasorEtAl2009}).  

\begin{figure}
   \centering
   \includegraphics[height=0.48\linewidth,angle=-90]{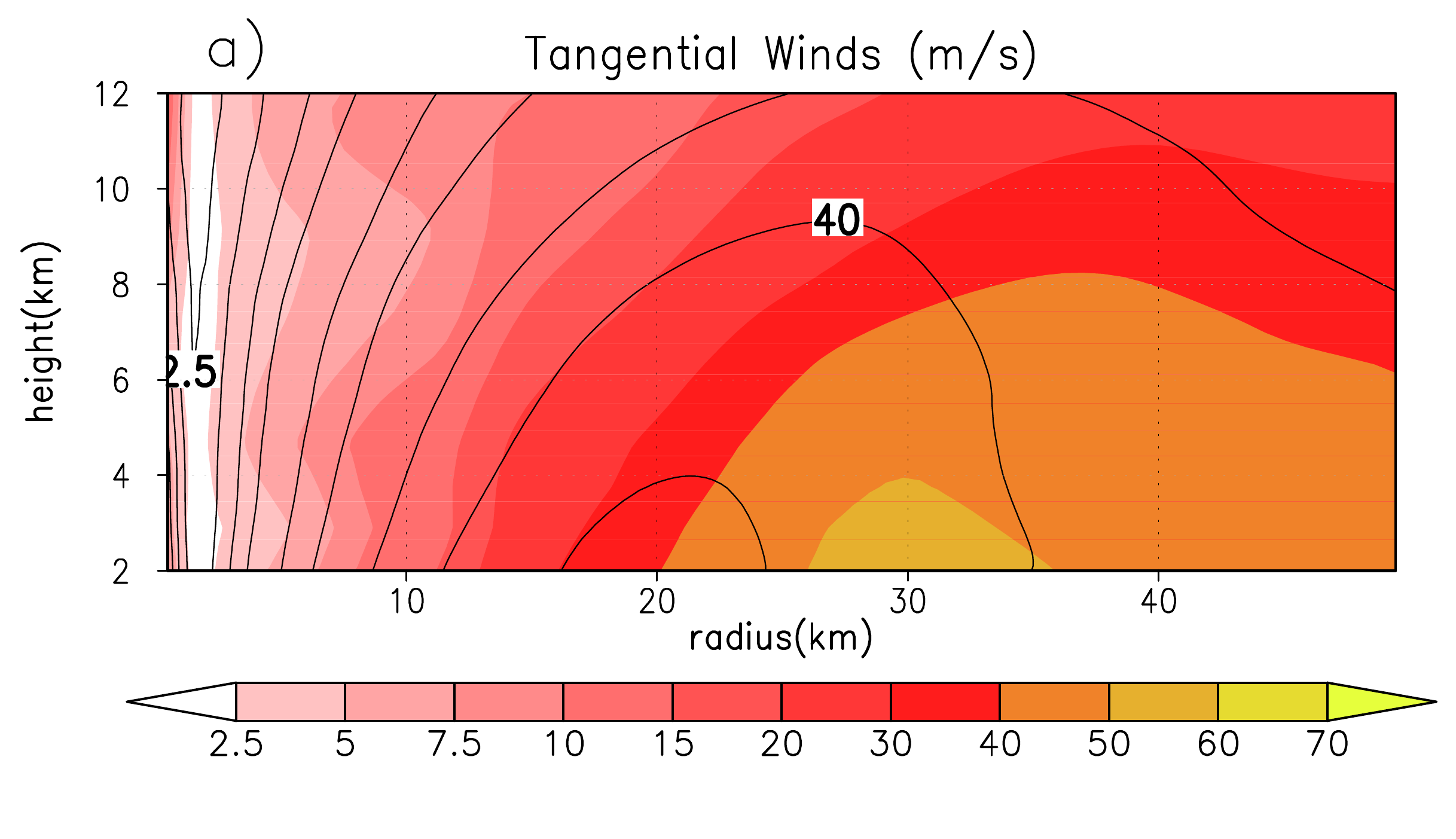}
   \includegraphics[height=0.48\linewidth,angle=-90]{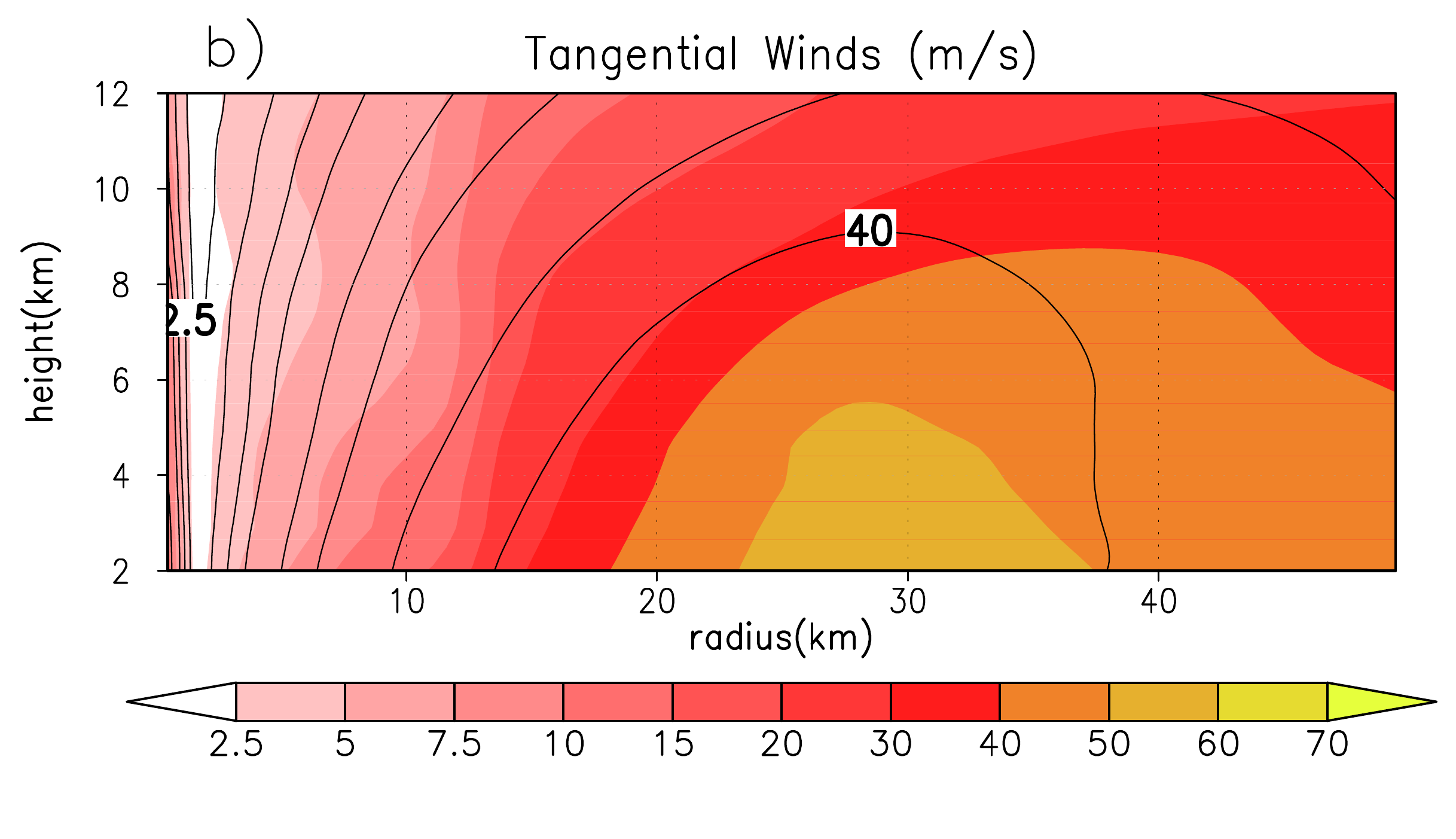}\\
   \includegraphics[height=0.48\linewidth,angle=-90]{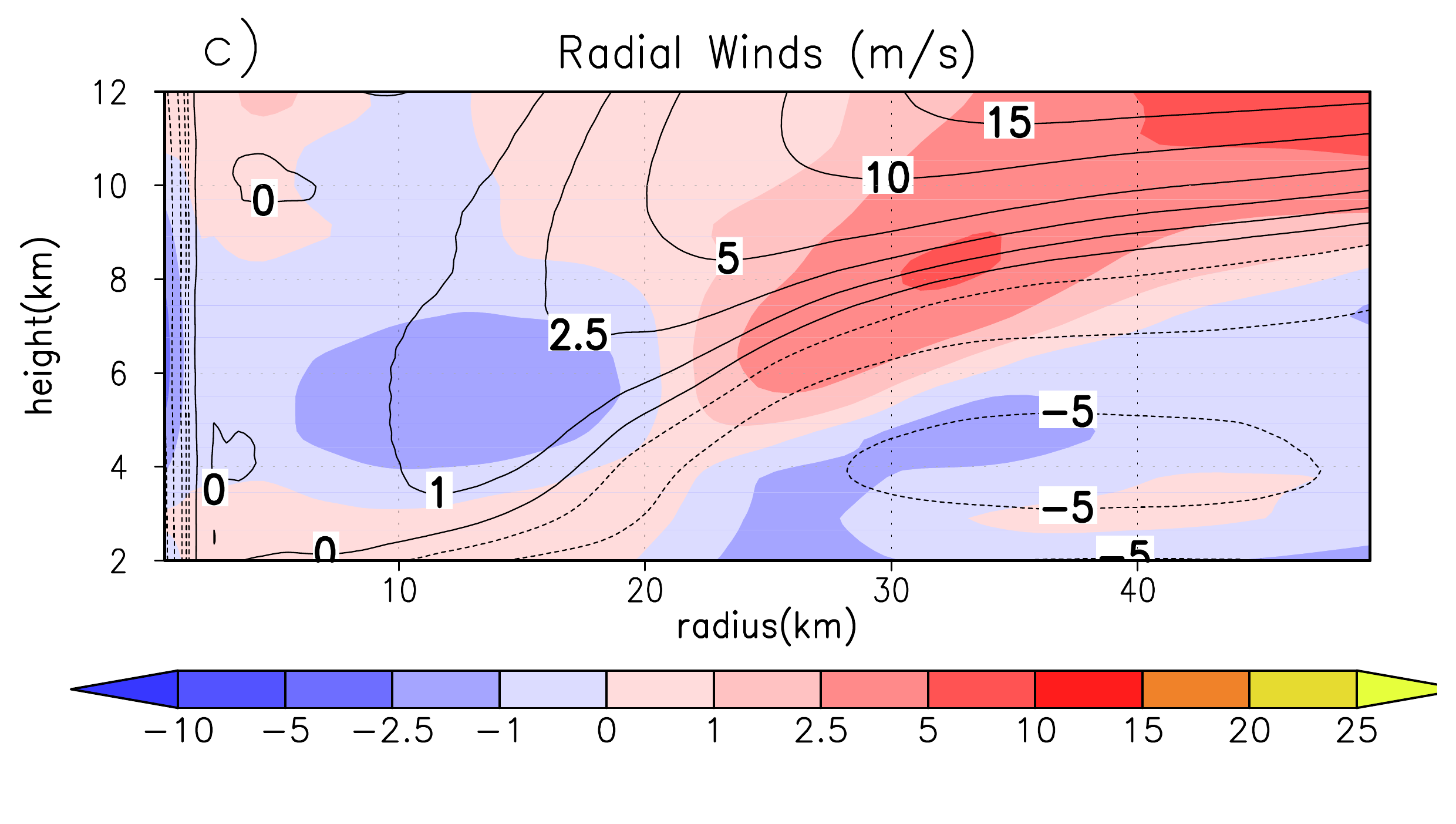}
   \includegraphics[height=0.48\linewidth,angle=-90]{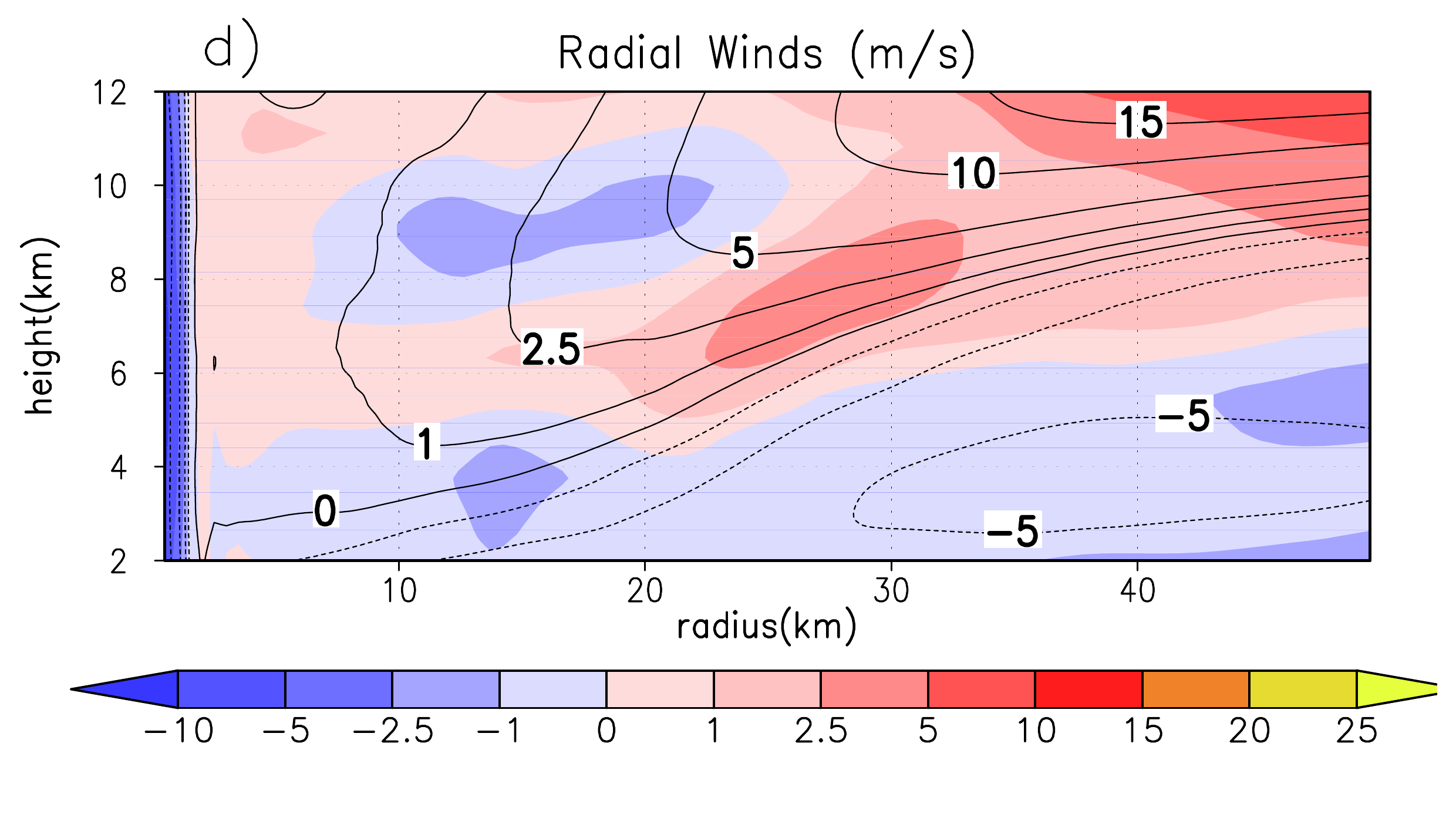}\\
   \includegraphics[height=0.48\linewidth,angle=-90]{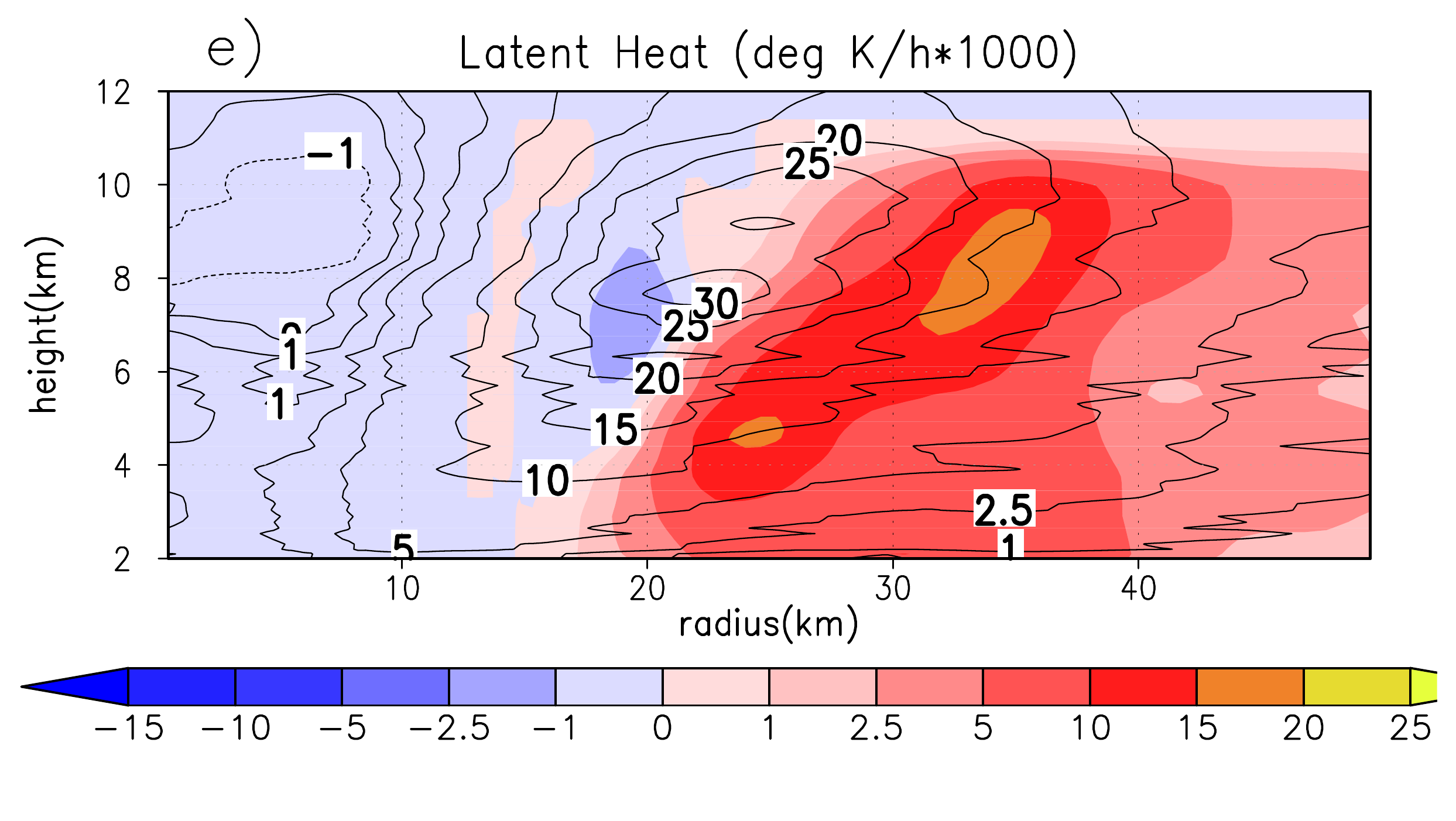}
   \includegraphics[height=0.48\linewidth,angle=-90]{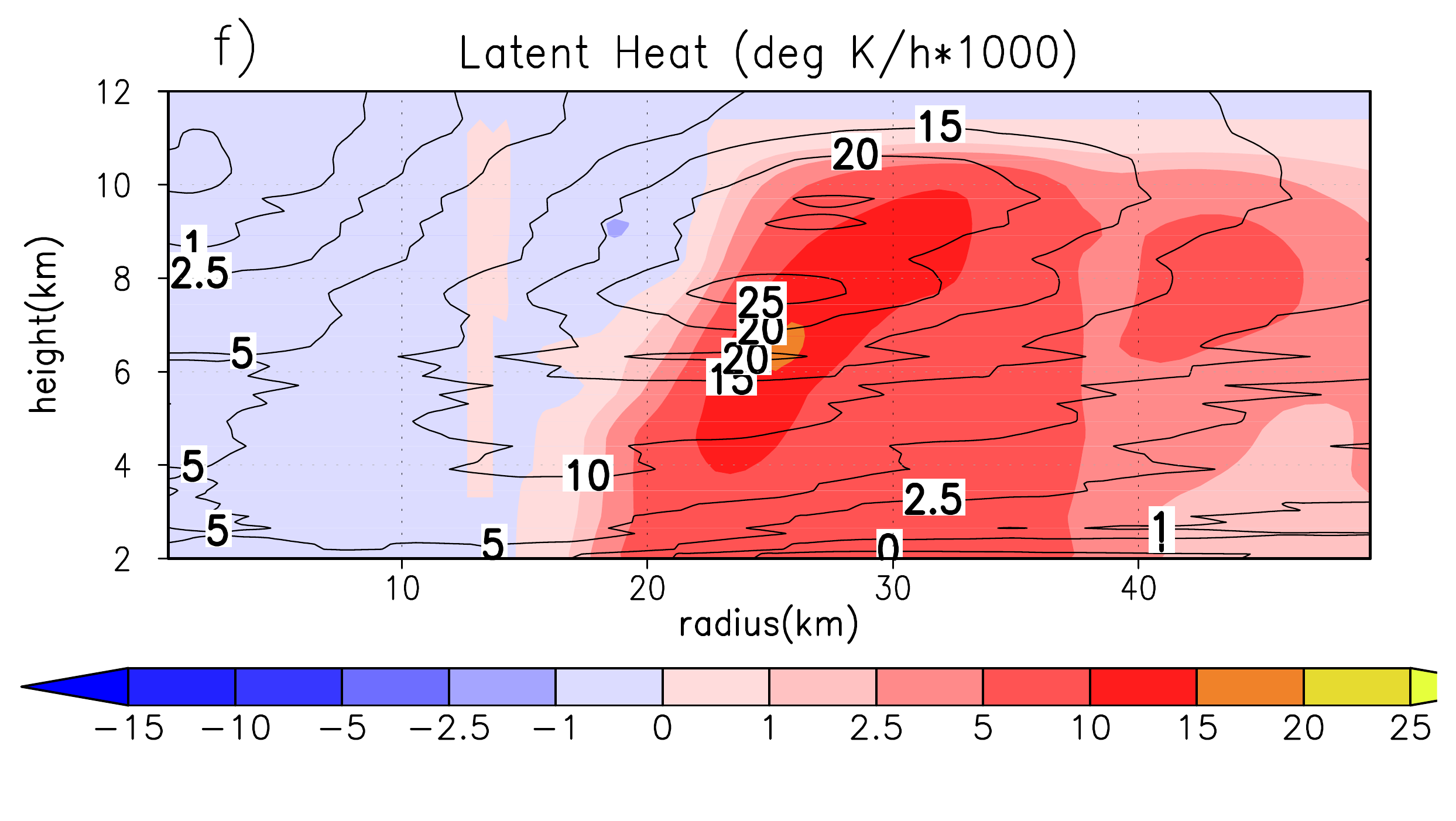}
   \caption{Comparisons between azimuthally-averaged profiles
    for the ensemble average (contours) and 
    observations (shaded) for tangential winds
    (top), radial winds (center), and latent heat (bottom). Time periods
    for comparisons are at flight leg 5 (2117 UTC) and 9 (2333 UTC).}
   \label{fig:ModelObs}
\end{figure}
\begin{figure}
    \centering
    \includegraphics[height=0.48\linewidth,angle=-90]{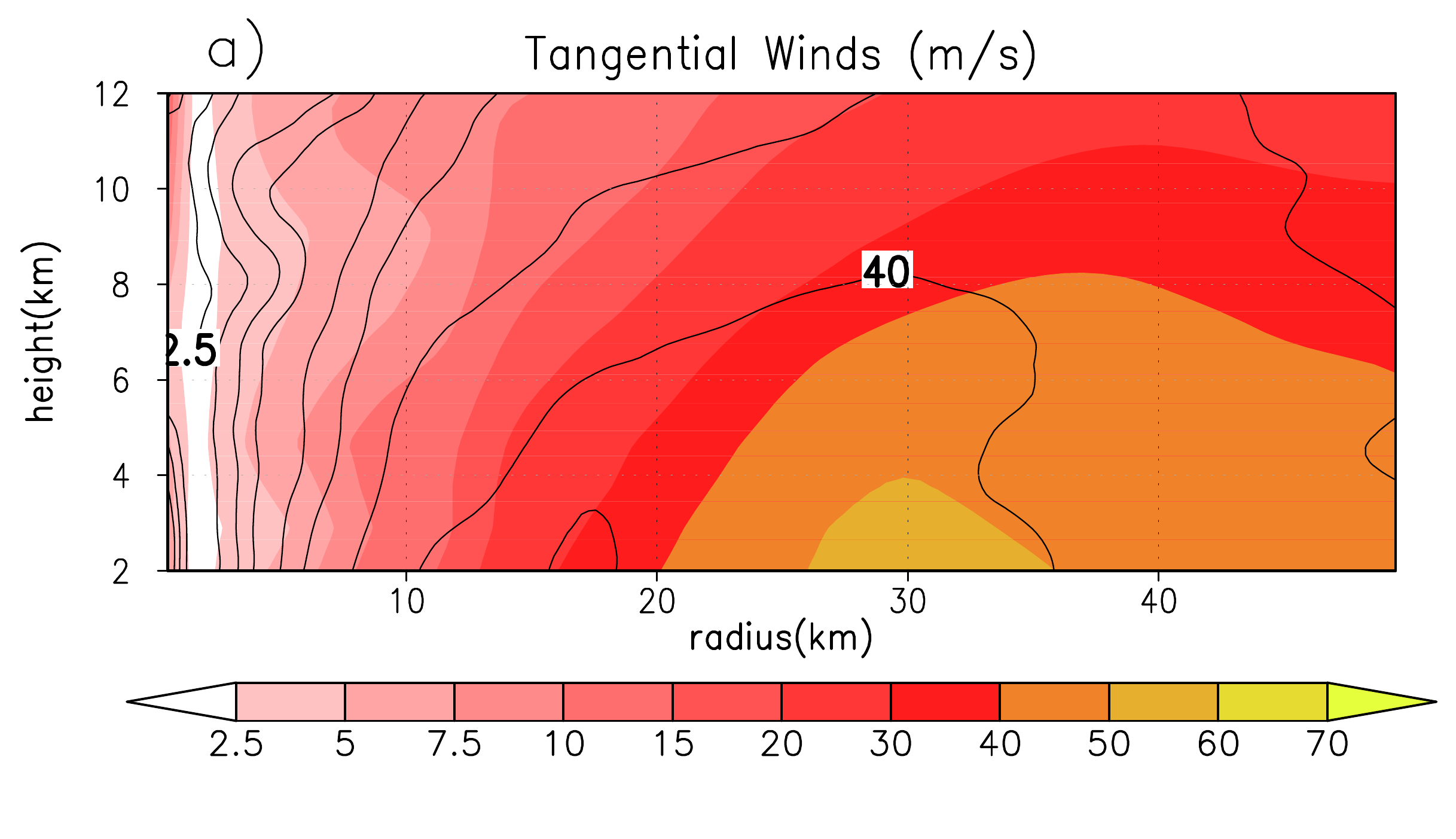}
    \includegraphics[height=0.48\linewidth,angle=-90]{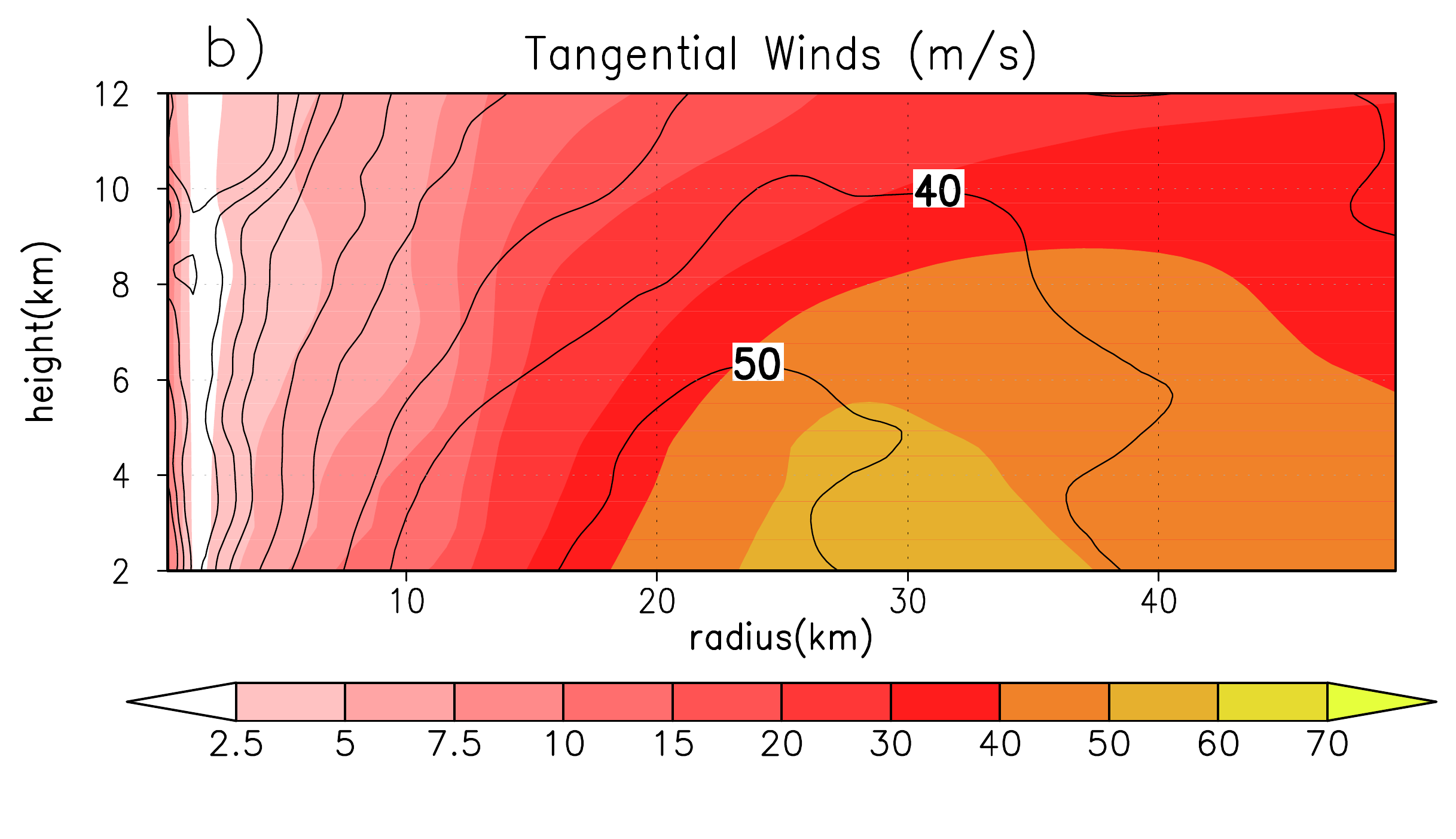}\\
    \includegraphics[height=0.48\linewidth,angle=-90]{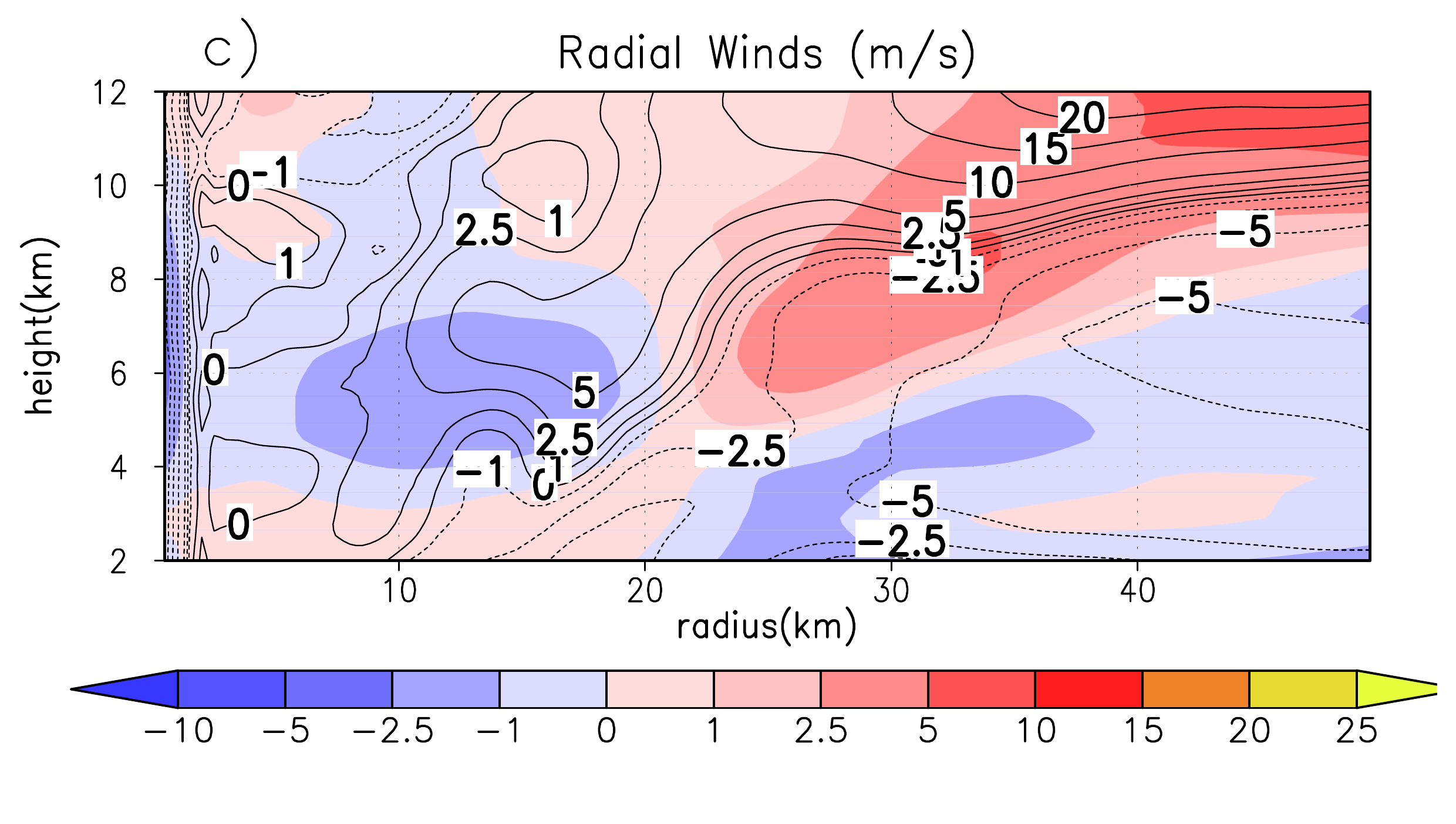}
    \includegraphics[height=0.48\linewidth,angle=-90]{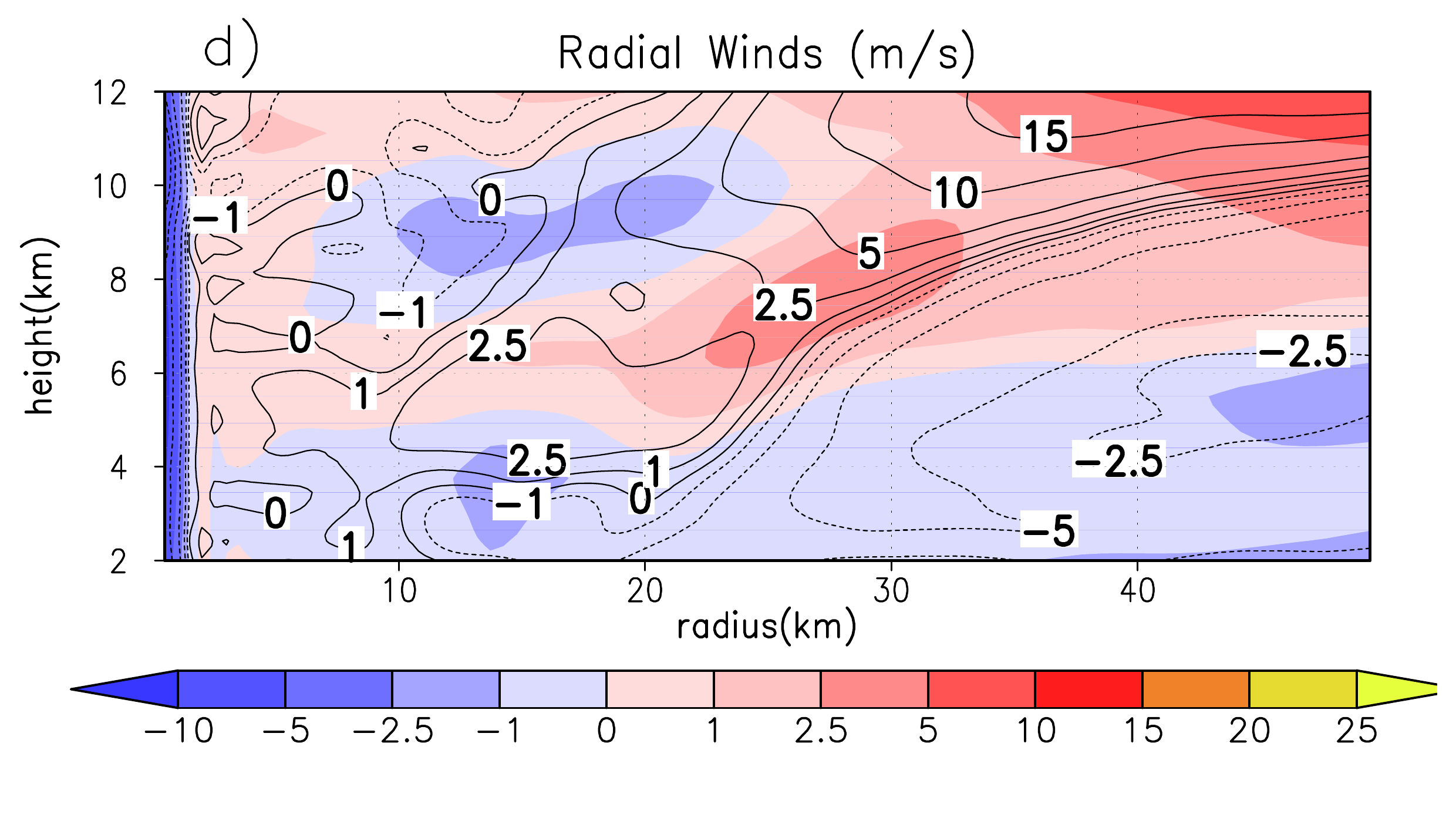}\\
    \includegraphics[height=0.48\linewidth,angle=-90]{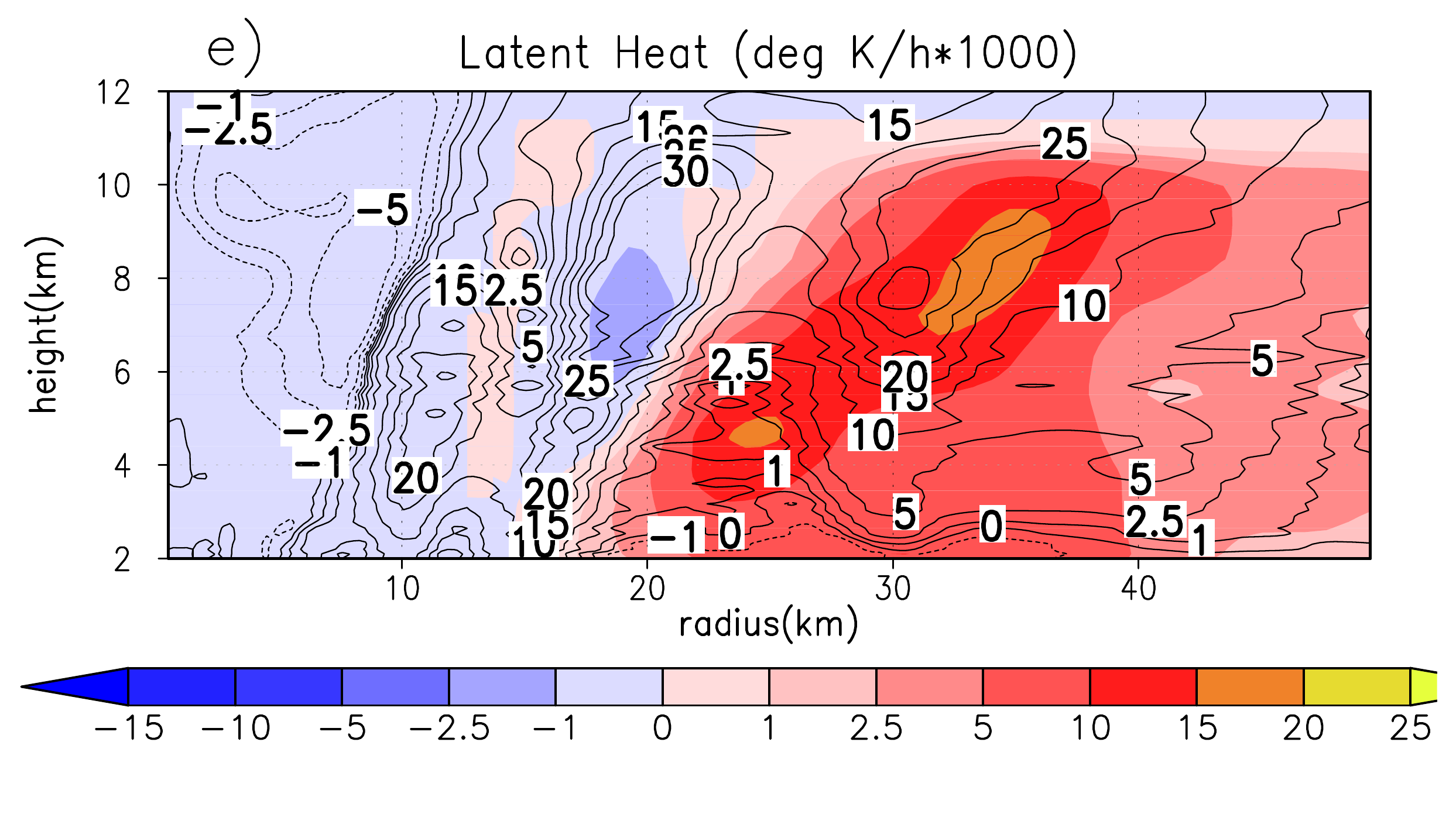}
    \includegraphics[height=0.48\linewidth,angle=-90]{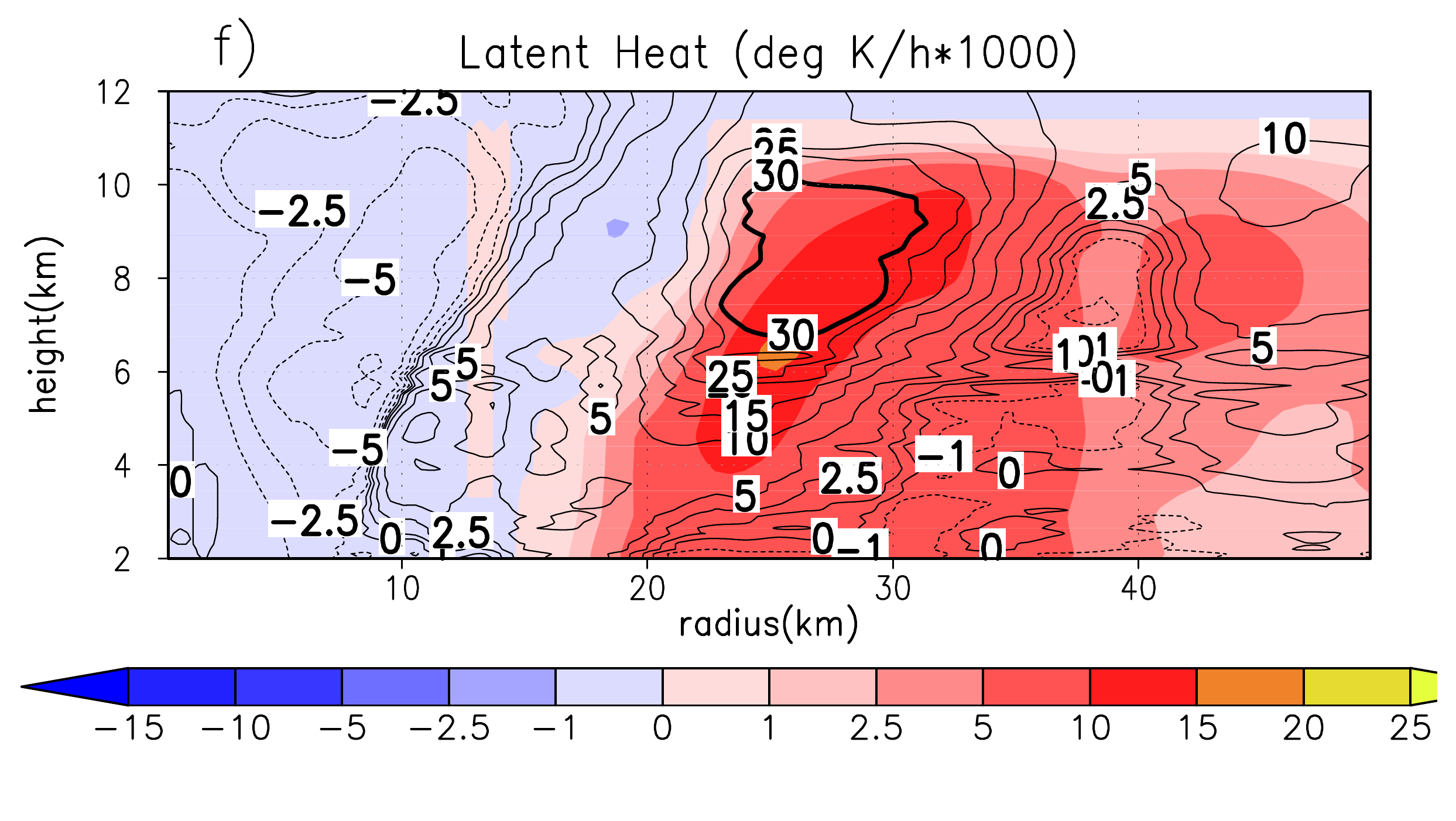}
    \caption{Comparisons between azimuthally-averaged profiles 
    for ensemble member 44 (contours) and
    observations  (shaded) for tangential winds
    (top), radial winds (center), and latent heat (bottom). Time periods
    for comparisons are the same as in the previous figure. }
    \label{fig:HG44}
\end{figure}
To provide another view of the simulated storm structure, both the
ensemble averaged and ensemble~44 azimuthal structures were compared
to observations and are presented for two flight legs in
Figs.~\ref{fig:ModelObs}-\ref{fig:HG44}.  Common disagreement with
observations can be seen in both figures: First the simulated radius
of maximum wind (RMW) and hence eyewall size is about $5-10$~km
smaller than in the observations. While the simulated ensemble
averaged azimuthal tangential component of the wind lies within $\sim
10$~m~s$^{-1}$ of the observations, more notable differences are seen
in the radial component of the wind with magnitudes rarely exceeding
$10$~m~s$^{-1}$ in the observations and the simulation consistently
producing magnitudes exceeding $15$~m~s$^{-1}$. Similar overestimation
is produced in the latent heat fields, with values exceeding $25$ or
even $30$ ($1000$~K~h$^{-1}$) in the simulation with the observations
showing values marginally reaching $20$ ($1000$~K~h$^{-1}$).

Despite these noteworthy differences, the HIGRAD model is able to
reproduce the slope/tilt of radial, tangential and latent heat fields
with a reasonable degree of realism. Moreover, the heights above sea
level of the contours encompassing the largest simulated values of
those three fields are in overall good agreement with observations.
Note, that the larger values of latent heat are required to compensate
for the impact of spurious evaporation (\cite{ReisnerJeffery2009}) common
to most cloud models with the consequences of this evaporation being
discussed later in this manuscript.  Additional plots were made for
the remaining 8 flight legs (not shown) and displayed similar
attributes.


\subsection{Twin-Experiments for Parameter Estimation}
\label{sec:TwinExperiments}

\begin{table}
   \centering
   \begin{tabular}{r|r}
      \hline\hline
      parameter & value \\
      \hline
      surface moisture & 9.325522e-02 \\
      wind shear & 4.968604e-01 \\
      turbulent length scale & 3.753693  \\
      surface friction & 1.443062 \\
      \hline
   \end{tabular}
   \caption{Parameter values used for the reference run, from where
   synthetic data is used for the twin-experiments.}
   \label{tab:RefParams}
\end{table}
To asses the reliability of
parameter estimation within the current context, and the amount of
observational data needed for the estimation, a series of
twin-experiments were performed.  A synthetic observational data set
is produced from a reference model run with the specific parameter
values give in Table~\ref{tab:RefParams}, and initialized according to
Section~\ref{sec:ModelSetupStructure}. These reference parameters
were selected near the ensemble average with white noise added to
them.


\begin{table}
   \centering
   \begin{tabular}{r|r}
      \hline\hline
      experiment & No. obs ($m$) \\
      \hline
      TE1 & 20 \\
      TE2 & 63 \\
      TE3 & 200 \\
      TE4 & 632 \\
      TE5 & 2000 \\
      TE6 & 6325 \\
      TE7 & 20000 \\
      TE8 & 63246 \\
      TE9 & 200000 \\
      \hline
   \end{tabular}
   \caption{Number of observations used in the twin-experiments for
   parameter estimation.}
   \label{tab:TwinExp}
\end{table}
\begin{figure}
   \centering
   \includegraphics[width=\linewidth]{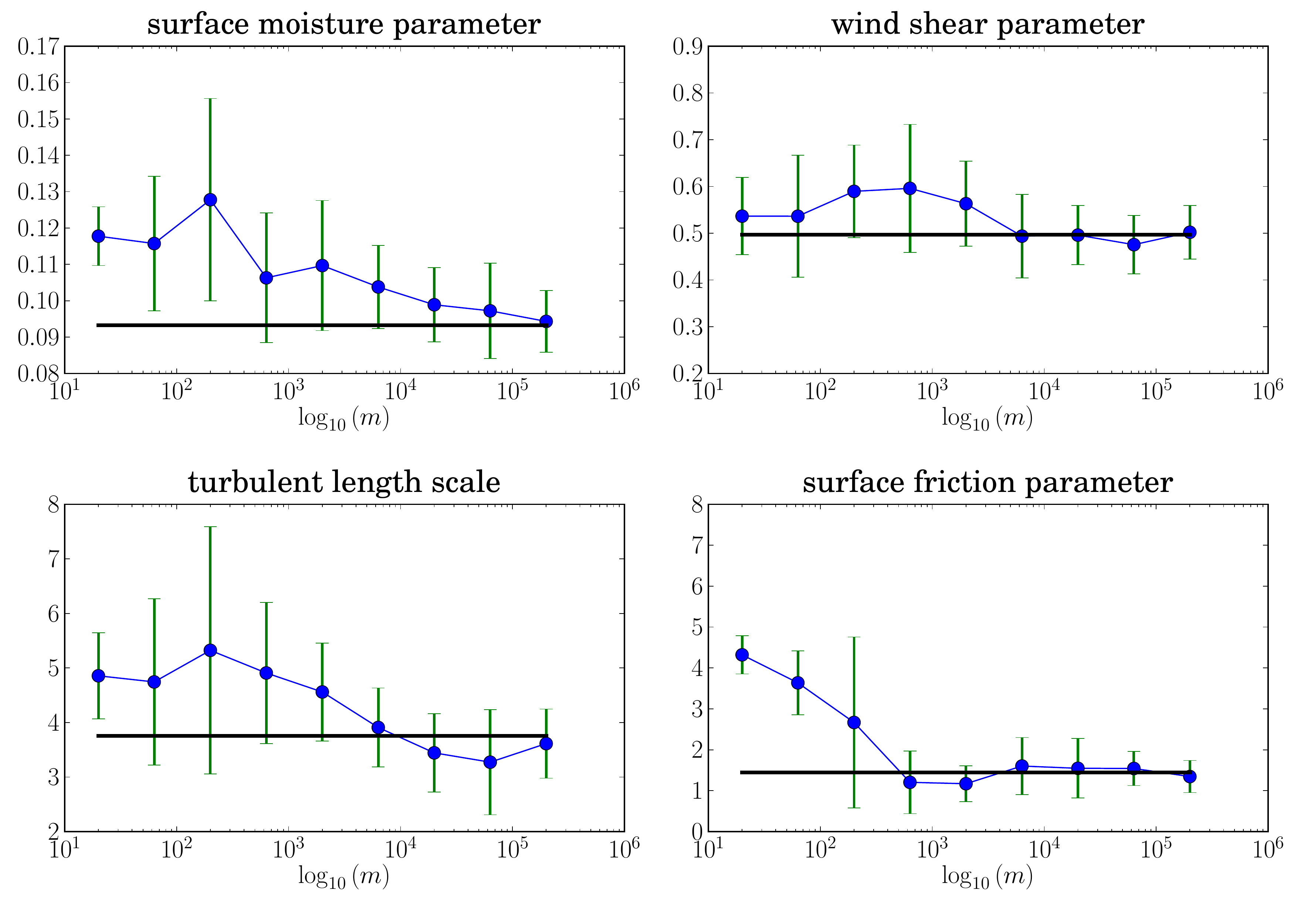}
   \caption{EnKF parameter estimation as a function of number of
   latent heat observations assimilated.  The latent heat observations
   were added in locations were the ensemble is most sensitive to
   changes in the parameters (Section \ref{sec:EnKFSetup}).}
   \label{fig:TwinParametersLatentHeat}
\end{figure}
In their paper \cite{TongXue2008} found that to simultaneously
estimate the model state and five parameters with the ensemble square
root filter, using their rank parameter-observation procedure, they
needed only 30 observational data points. Given that our parameter
estimation setting is significantly different from their approach, it
is not entirely obvious whether only small amounts of data are
required to conduct the parameter estimation or, in the other limiting
situation, more data than is currently available is needed for
undertaking the estimates. To address this issue, nine parameter
estimation experiments, named TE1 to TE9, were performed for different
amounts of data, given in Table~\ref{tab:TwinExp}. The data being used
for parameter estimation is the latent heat field from the reference
run, where the data is selected according to the procedure described
in Section~\ref{sec:EnKFSetup}. The first observation set was taken at
$t = 6$~hours of simulation time, afterwards nine more observation
sets were taken at 30 minute intervals, which makes a total of 10
observational data set over a five hour window. For each experiment,
the parameters are estimated according to
Section~\ref{sec:ConsiderationsEnKF}, that is, parameter estimates for
the ensemble are computed with the EnKF at each observational time,
and then a final parameter estimate is then computed by averaging over
ensemble and then time, as in equation \eqref{eq:ParameterEstimate}.
Figure~\ref{fig:TwinParametersLatentHeat} shows the parameter
estimates for each experiment, where the vertical lines indicate the
time variance of the parameter estimate. The figure clearly shows the
impact of the additional data on the parameter estimates with a
noticeable reduction in the error for all for parameters. Furthermore,
it is only when approximately 200,000 observations are used that all
four parameters converge to the correct values. This is highly
relevant since it indicates that a significant amount of data is
required, in this context, to correctly estimate the values of the
parameters. A similar result was obtained when the horizontal wind
field or reflectivity were used to estimate the parameters.

\subsection{Parameter Estimation Experiments with Hurricane Guillermo
Observations}
\label{sed:ParamHurricaneGuillermo}

Given the large ensemble spread in various model fields, such as
intensity, and the ability of the model to reproduce observed data
suggests that estimation of the four model parameters is not only
possible with the EnKF, but should produce parameter estimates that
hopefully reduce model forecast errors. 

The number of observations used for parameter estimation, using the
EnKF in all subsequent experiment, were those identified with the
highest ensemble sensitivity located at 200,000 model grid-points (see
selection of observations in Section \ref{sec:EnKFSetup}).  The
assimilation is performed over latent heat (DA1), horizontal winds
(DA2), or both fields (DA3) started six hours into the ensemble
(corresponding to 1900~UTC).  With the inclusion of DA3, an assessment
regarding how the EnKF procedure weights two different observational
data sets and model results can be made and analyzed. Hence, this
section will not only highlight how the various parameter estimates
change when using different observational fields, but also how these
estimates change in time.

\begin{figure}
   \centering
   \includegraphics[width=\linewidth]{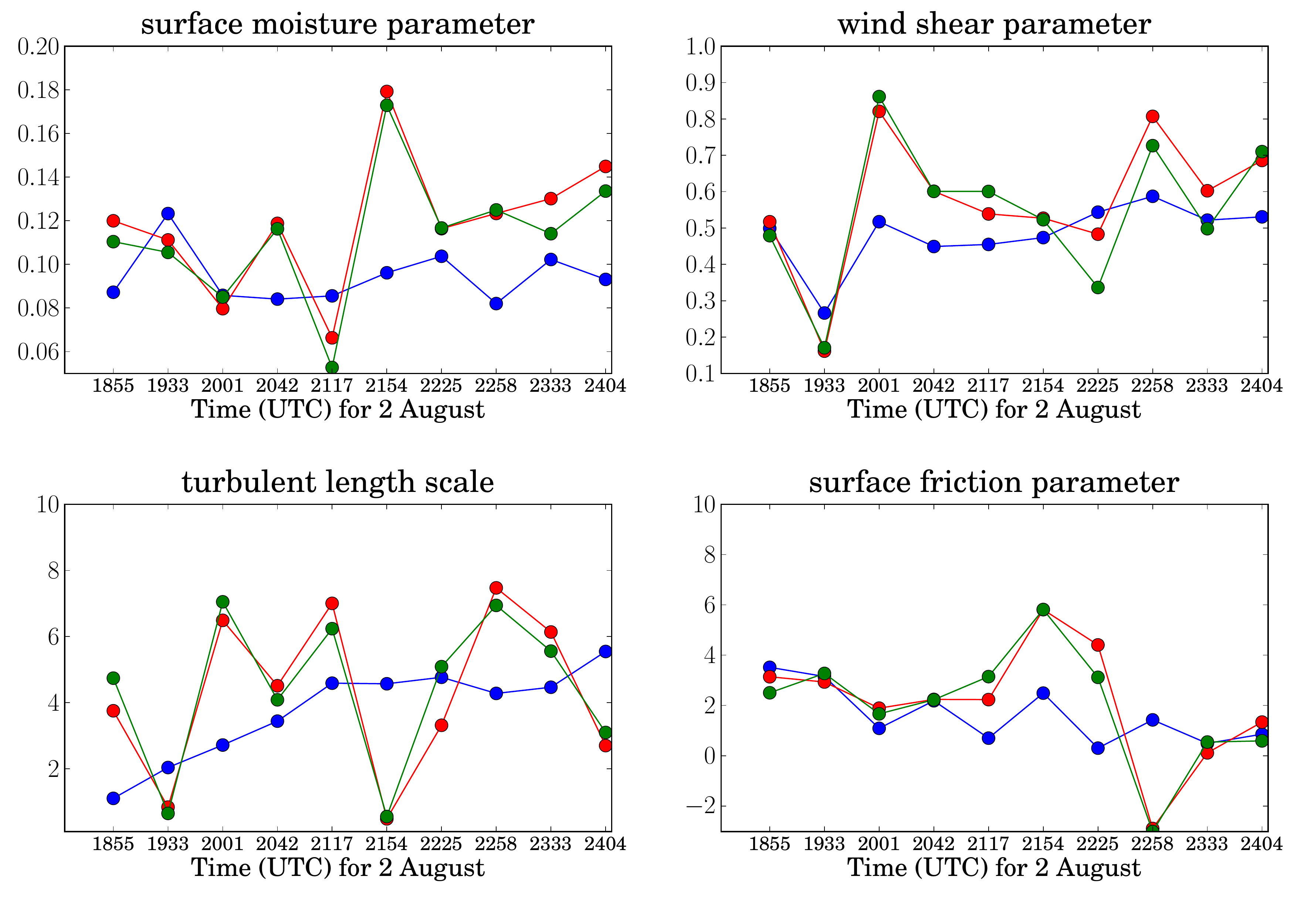}
   \caption{Time distribution of the ensemble average parameter
   estimates with EnKF from DA1 (blue line, latent heat), DA2 (red
   line, horizontal winds), and DA3 (green line, both latent heat and
   horizontal winds).} 
   \label{fig:AnalysisParamterTime}
\end{figure}
Figure~\ref{fig:AnalysisParamterTime} shows the time distribution of
the ensemble average parameter estimates with EnKF using the 10 data
time periods for DA1, DA2, and DA3. The largest temporal oscillations
in the parameter estimates are associated with DA2 and suggest that
the wind fields produced by the ensemble tend to oscillate more in
time than the latent heat fields. Likewise, the surface moisture and
the wind shear parameter estimates do not appear to change
significantly in time, whereas the turbulent length scale and surface
friction estimates either increase or decrease with time. Note, the
temporal changes in these two parameters could be due to numerical
errors and/or the impact of initial condition errors. For example,
Fig.~\ref{fig:ModelObs} shows that with time the areas of maximum
latent heating and/or winds from the ensemble are expanding outward
and hence this outward expansion, probably the result of numerical
diffusion, could explain the temporal changes in these two parameter
estimates.

\begin{table}[t]
   \centering
   \begin{tabular}{l|cccc}
      \hline\hline
      parameter/simulation or DA & HG 44 & DA1 & DA2 & DA3 \\
      \hline
      surface moisture & 1.944818e-01 &  9.431537e-02 & 1.189942e-01 & 1.132110e-01 \\
      wind shear & 8.122108e-01 &  4.843292e-01 &  5.744430e-01& 5.506012e-01 \\
      turbulent length scale & 4.524457 &  3.753072 & 4.271998 &  4.401512 \\
      surface friction & 2.005939 &  1.619931 & 2.120377 & 1.986076 \\
      \hline
   \end{tabular}
   \caption{Time average parameter values for each of the experiments
   DA1-DA3, and the parameter values for ensemble member 44 (HG 44).}
   \label{tab:ParameterTimeAverage}
\end{table}
\begin{figure}
    \centering
    \includegraphics[width=0.9\linewidth]{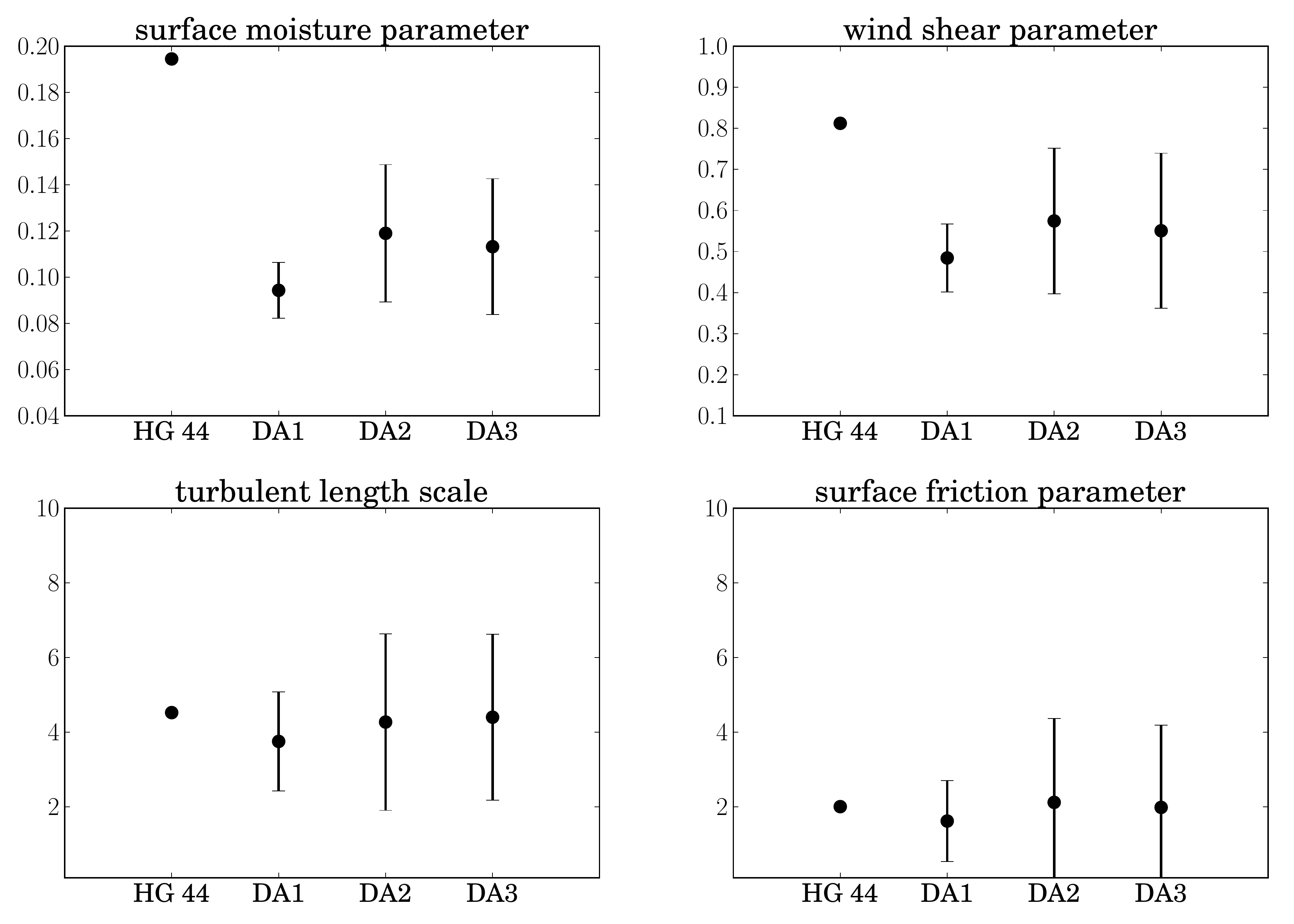}
    \caption{Analysis parameters averaged over time for ensemble
    member 44 (HG 44), DA1, DA2, and DA3. The vertical lines from the
    dots indicate the time variance of the parameter estimates each
    experiment.}
    \label{fig:AnalysisParameterAverage}
\end{figure}
Upon averaging the parameter estimates over the ensemble and in time,
differences between the various parameter values are small; however,
as will be shown in the next section these small differences in the
parameters do lead to rather significant differences in both the
structure and intensity of the simulated hurricanes. 
%
%
Table~\ref{tab:ParameterTimeAverage} show the time average parameter
values from the assimilation experiments, as well as the parameter
values used for ensemble member 44 prior to assimilation.
Fig.~\ref{fig:AnalysisParameterAverage} also reveals the various
parameter estimates are different than the parameter values used in
ensemble member 44, suggesting the importance of using a technique
such as the EnKF to obtain the estimates, instead of simply using
estimates obtained from a simulation that matches an observable such
as minimum sea level pressure. This ability of the EnKF to reasonably
fit the parameter values to the chosen observational data set is as
well illustrated by the time averaged estimates from DA3 that, as
expected, lie somewhere between the parameter estimates from DA1 and
DA2, except for the turbulent length scale parameter.

\subsection{Parameter error assessment}
\begin{figure}
   \centering
   \includegraphics[height=0.48\linewidth,angle=-90]{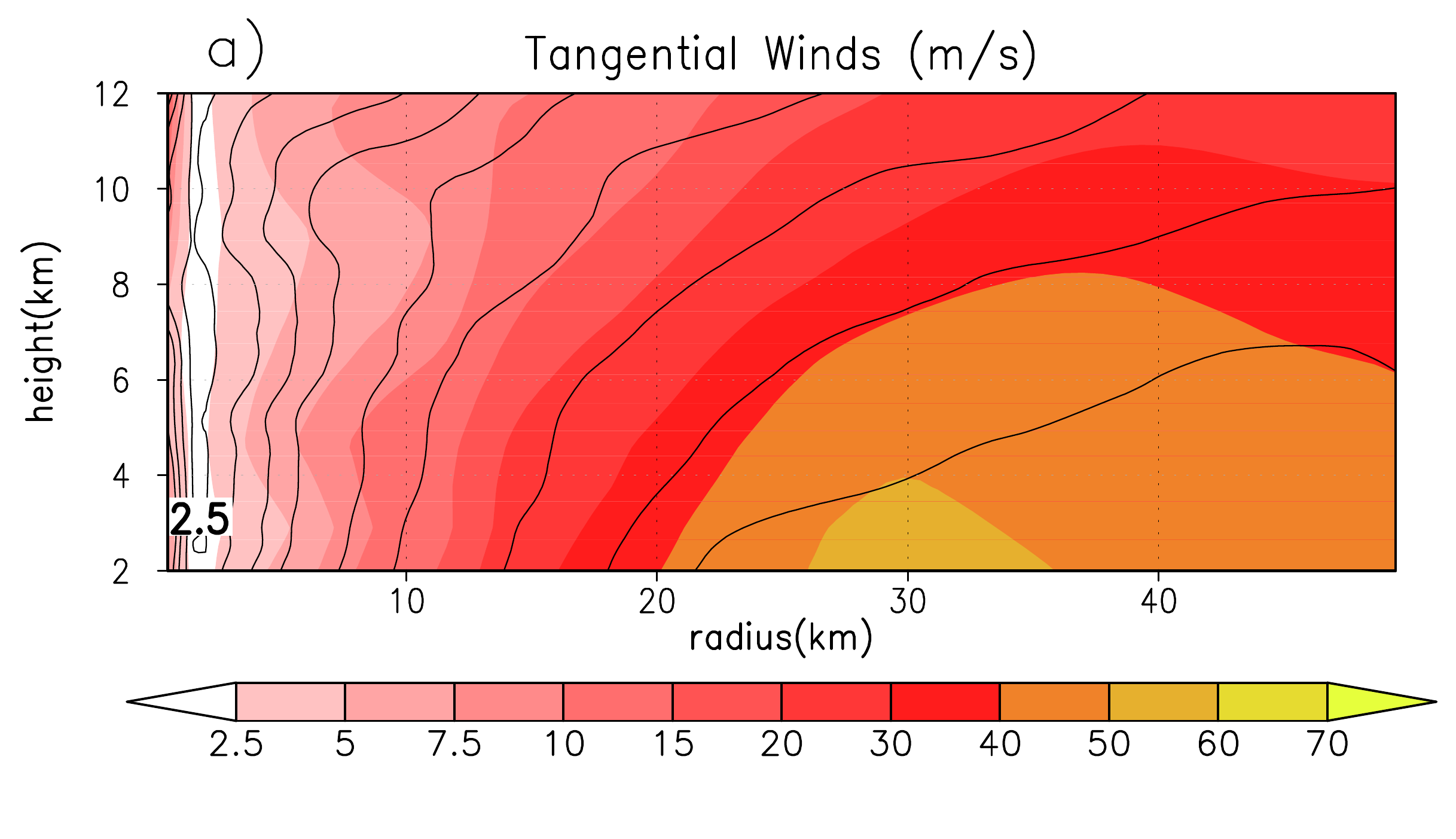}
   \includegraphics[height=0.48\linewidth,angle=-90]{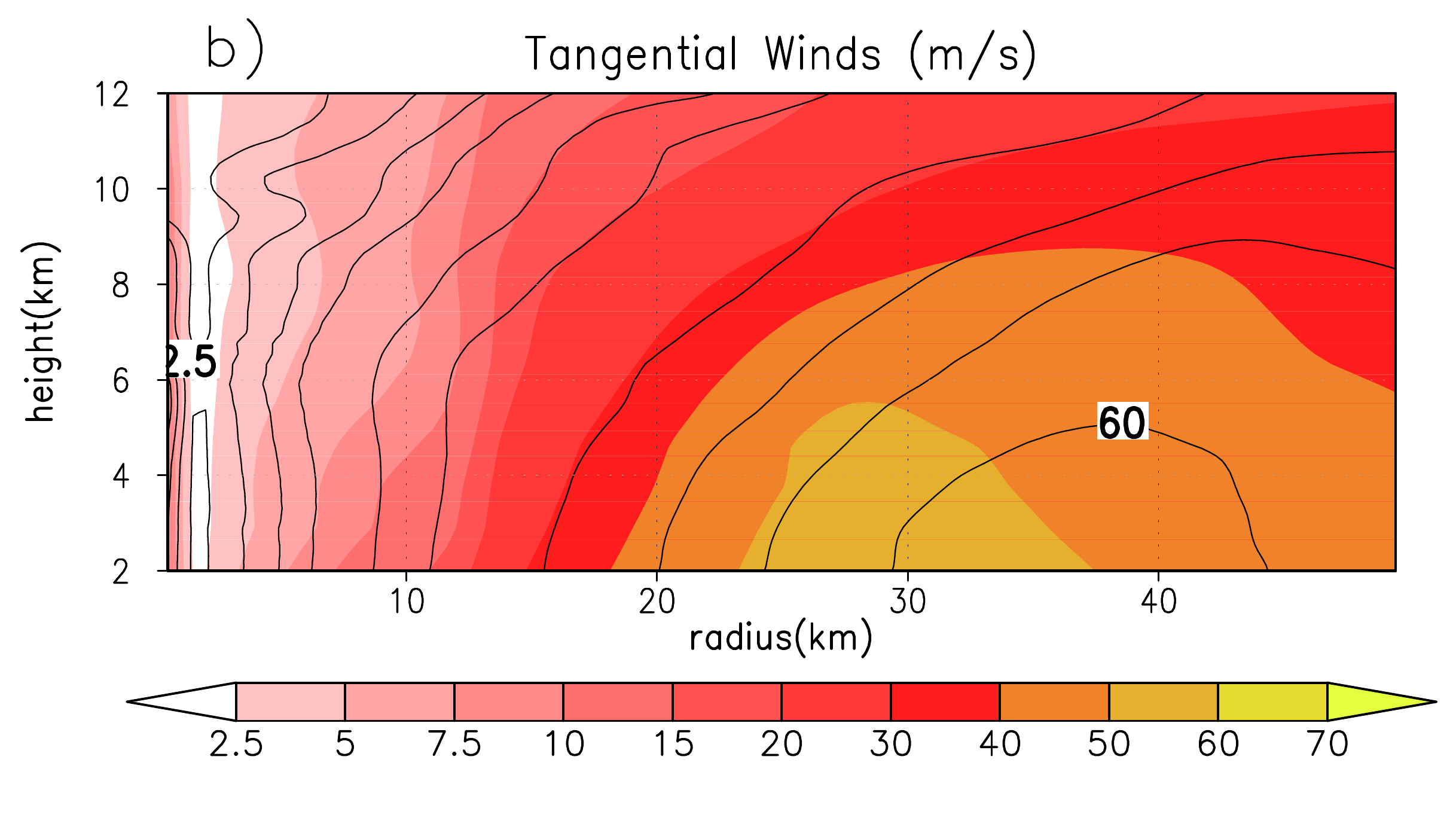}\\
   \includegraphics[height=0.48\linewidth,angle=-90]{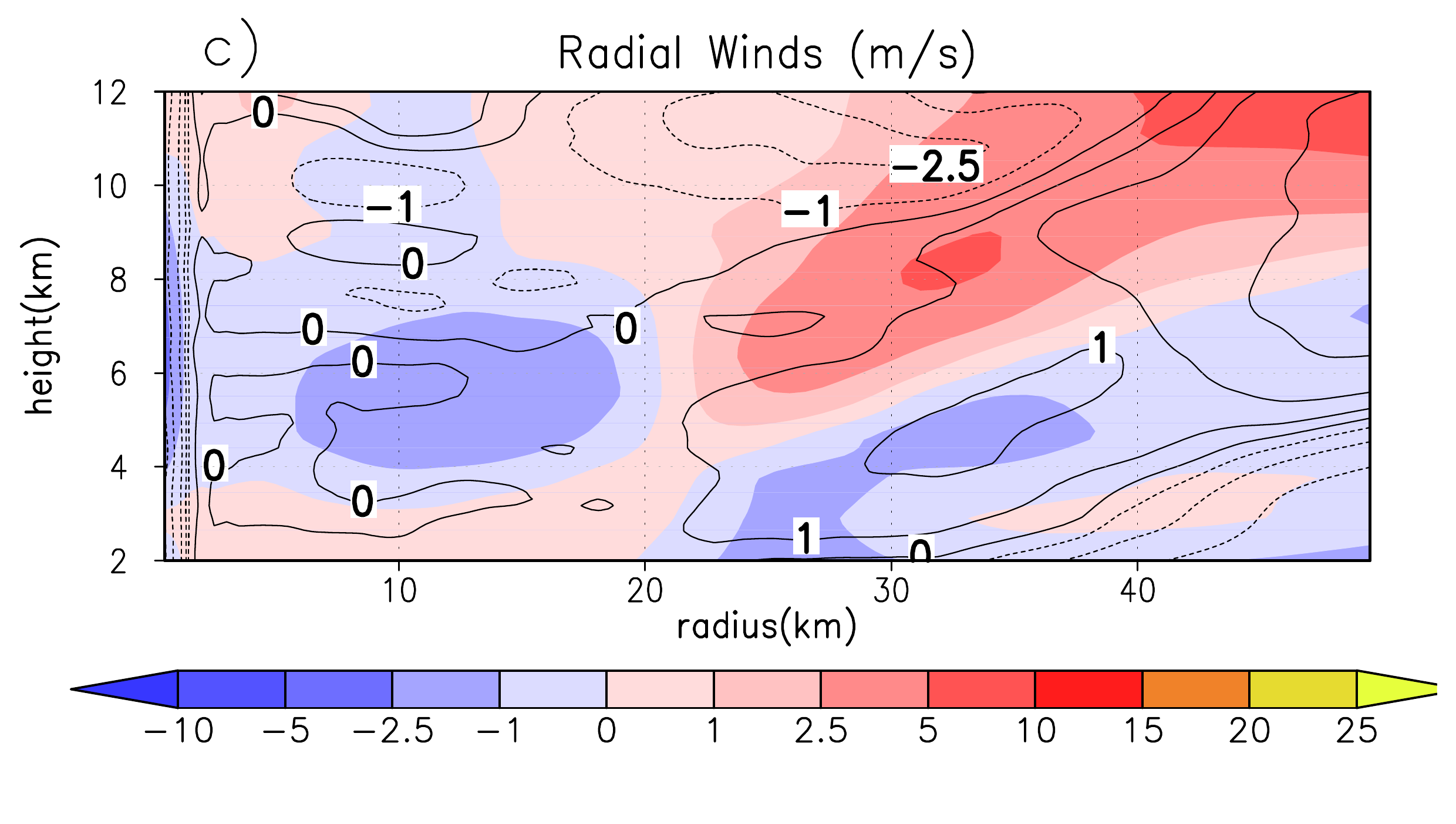}
   \includegraphics[height=0.48\linewidth,angle=-90]{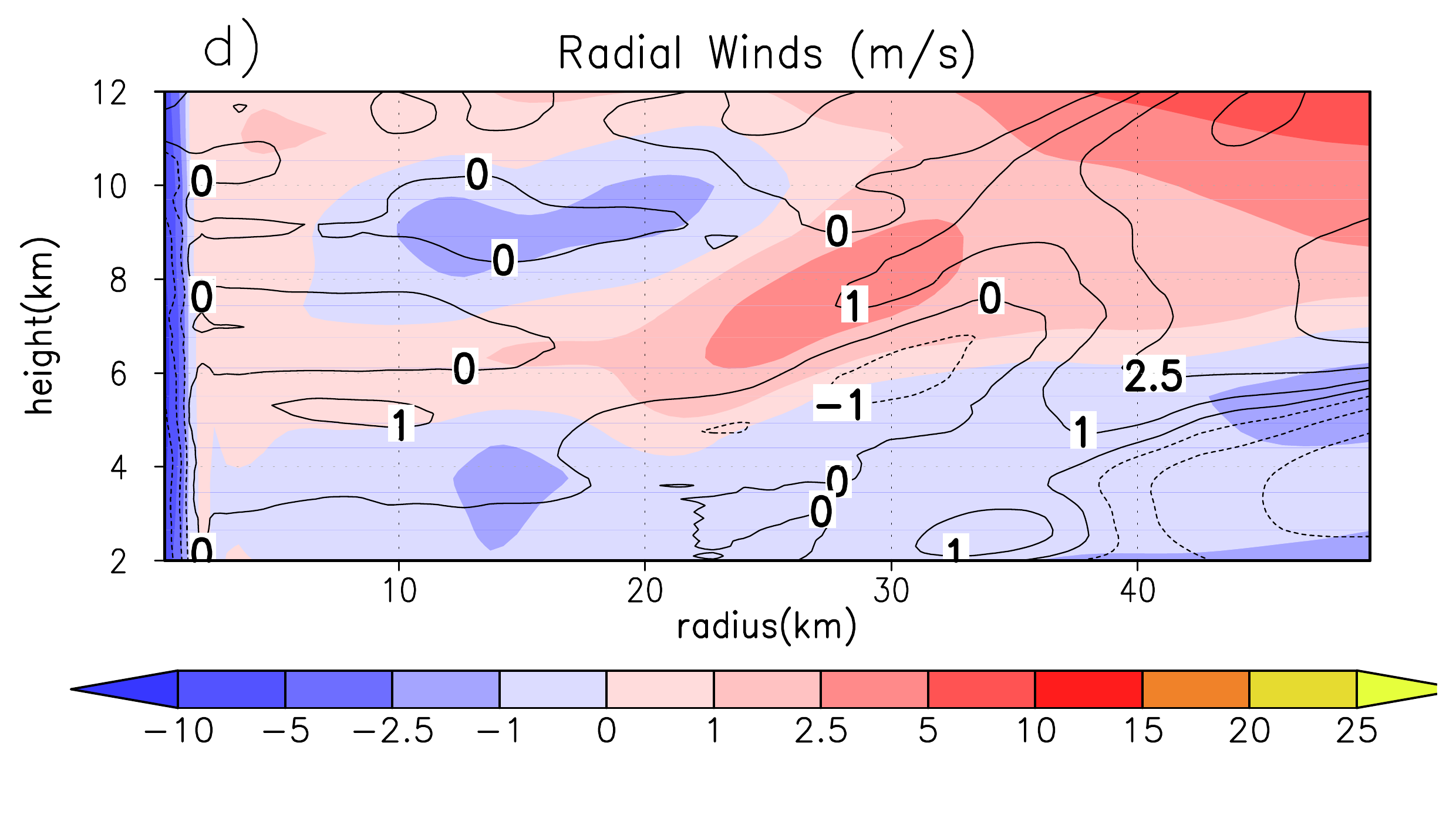}\\
   \includegraphics[height=0.48\linewidth,angle=-90]{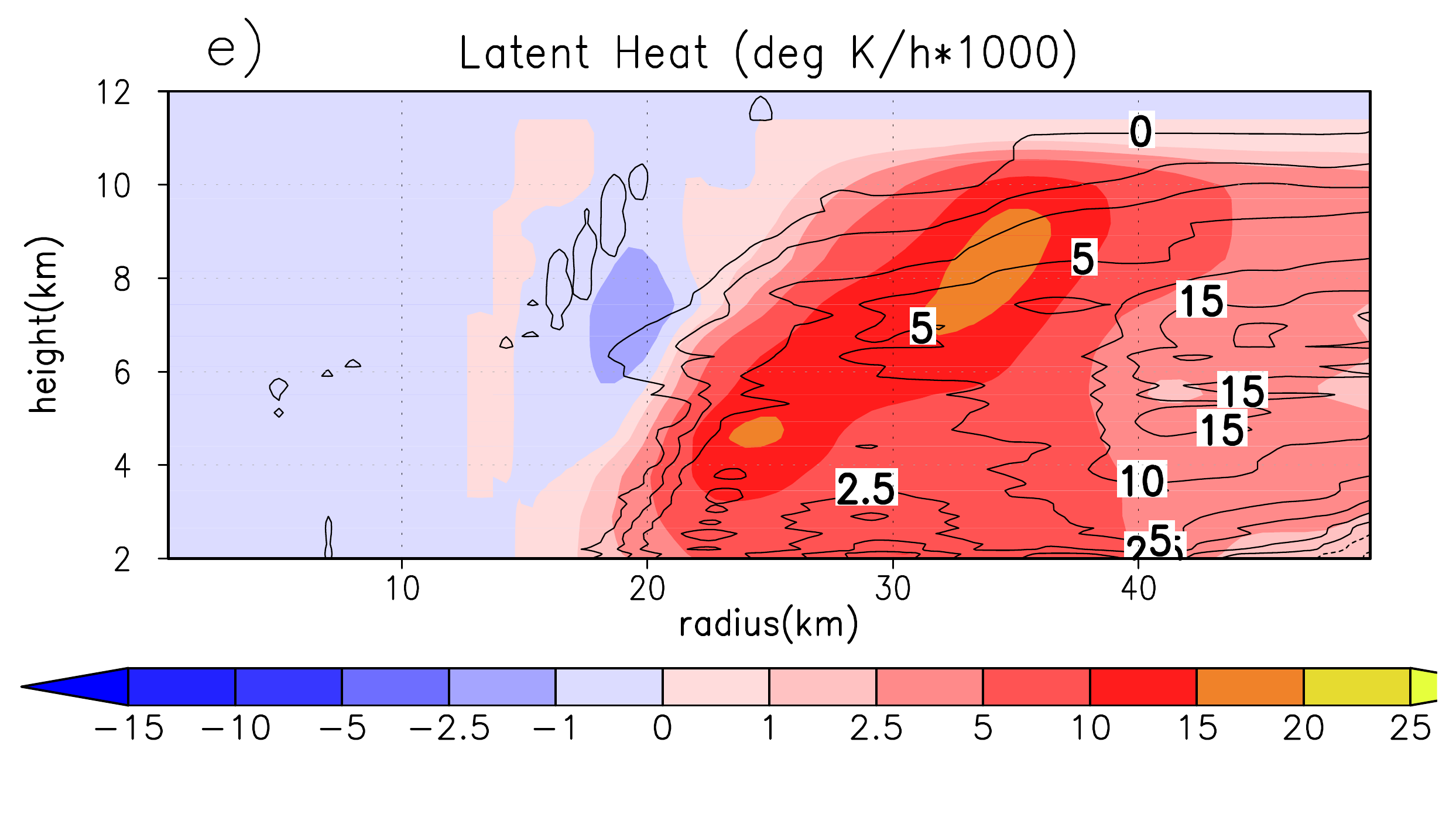}
   \includegraphics[height=0.48\linewidth,angle=-90]{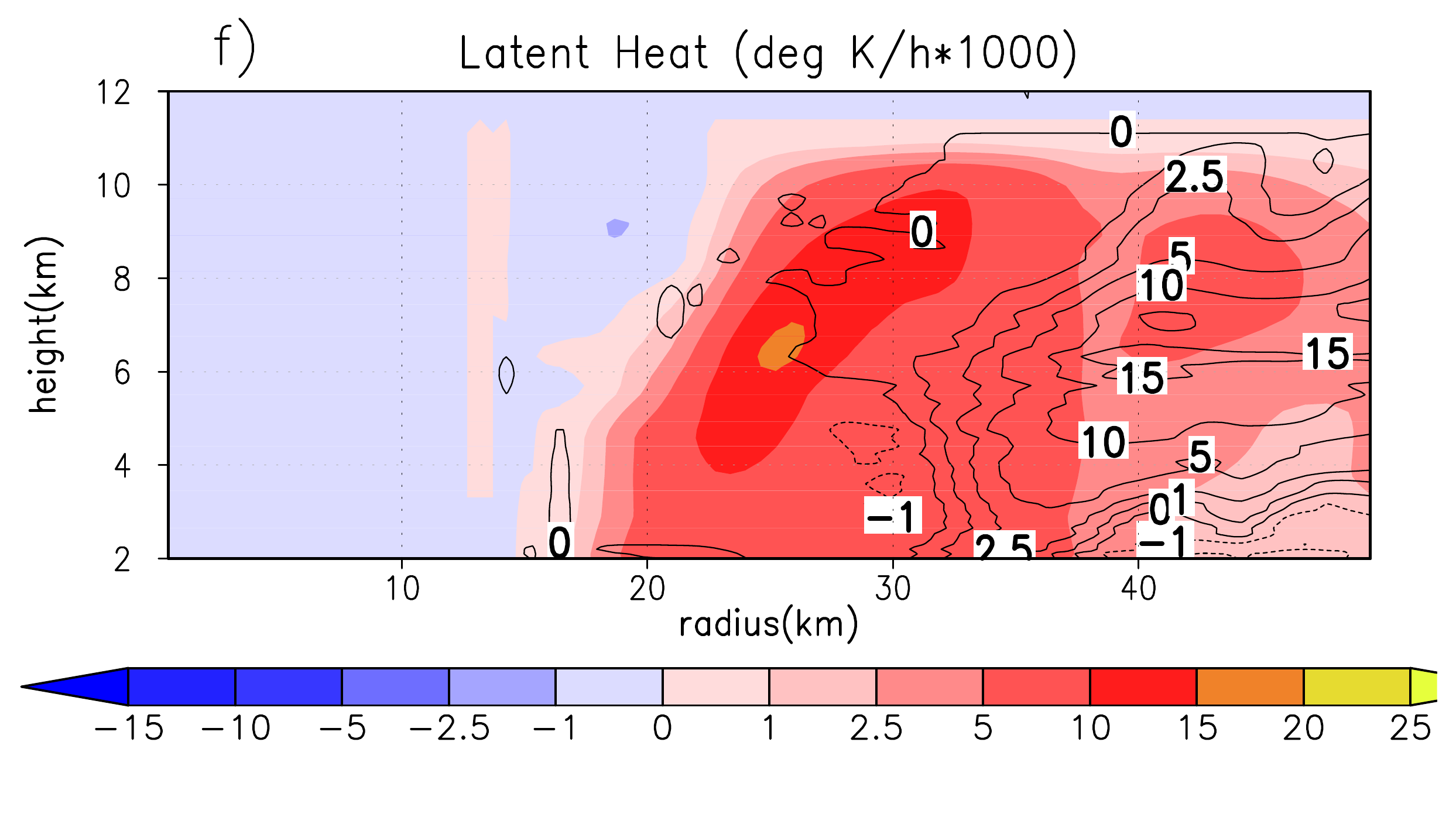}
   \caption{ Comparisons of the azimuthally-averaged profiles between
   a model simulation (contours) using estimated parameters
   from DA1 and observations (color shaded). Plots for
   tangential winds (top), radial winds (center), and latent heat (bottom)
   for flight leg 5 (2117 UTC) and 9 (2333 UTC).}
   \label{fig:ModelObsLH}
\end{figure}
\begin{figure}
   \centering
   \includegraphics[height=0.48\linewidth,angle=-90]{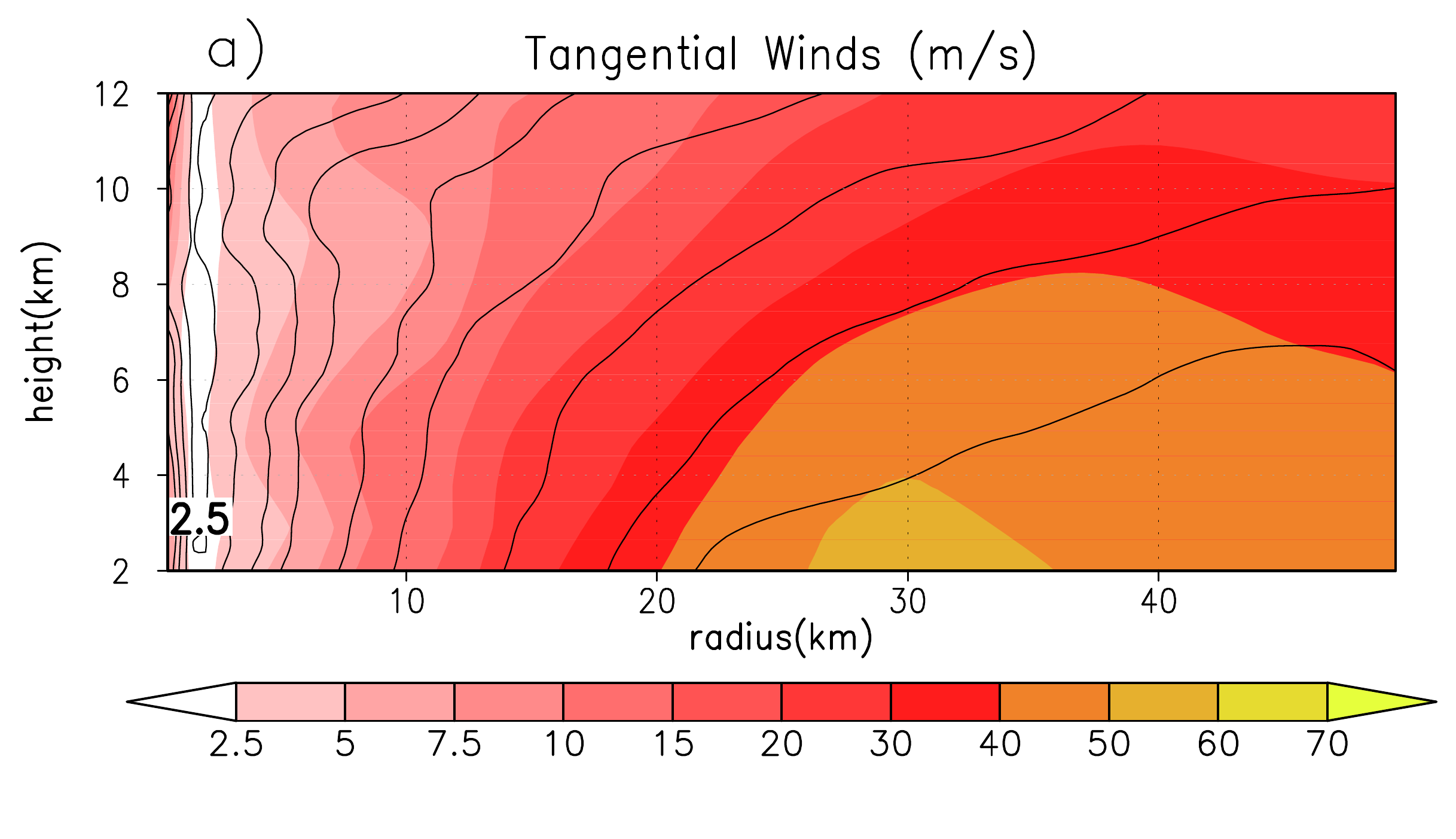}
   \includegraphics[height=0.48\linewidth,angle=-90]{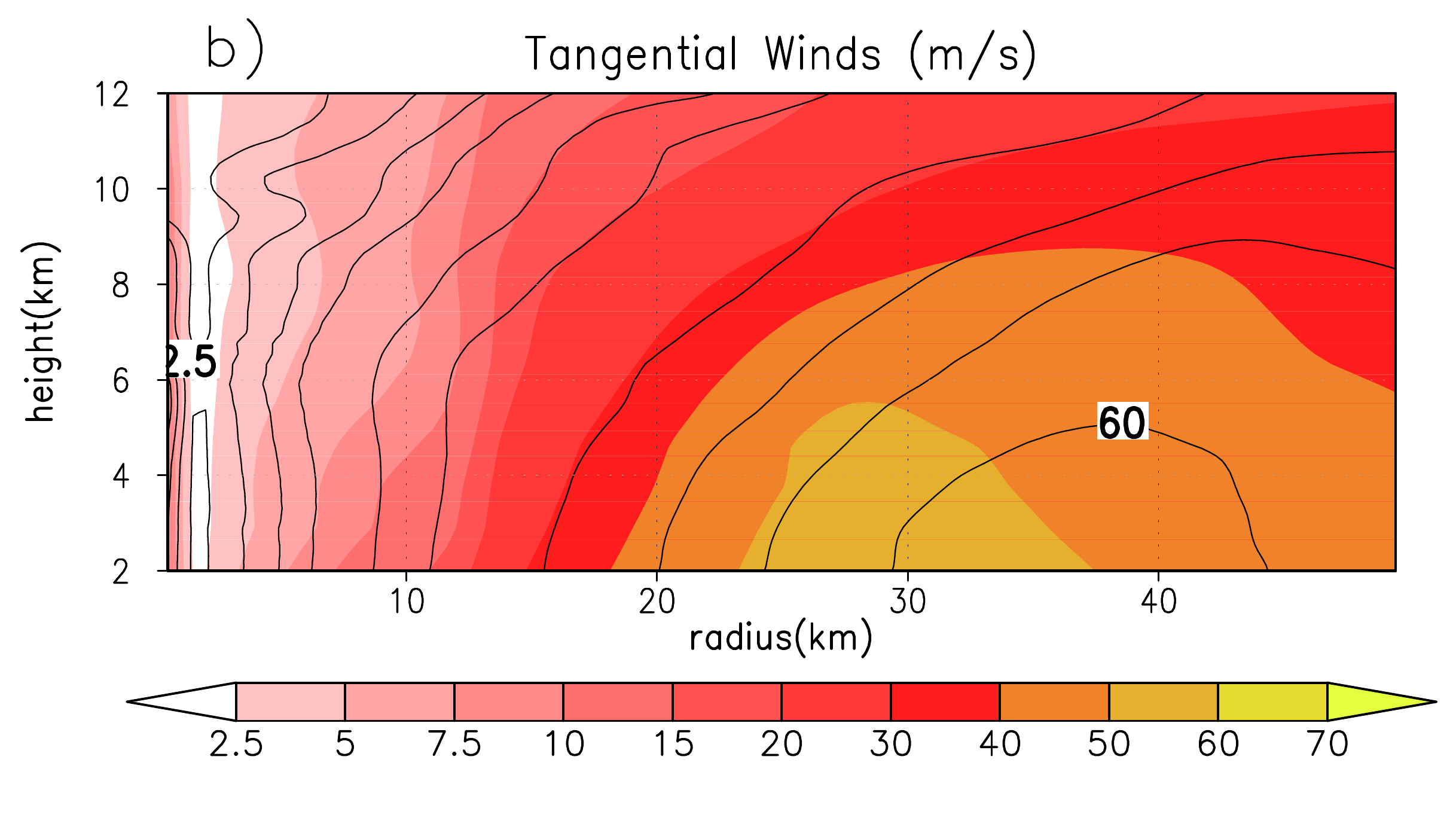}\\
   \includegraphics[height=0.48\linewidth,angle=-90]{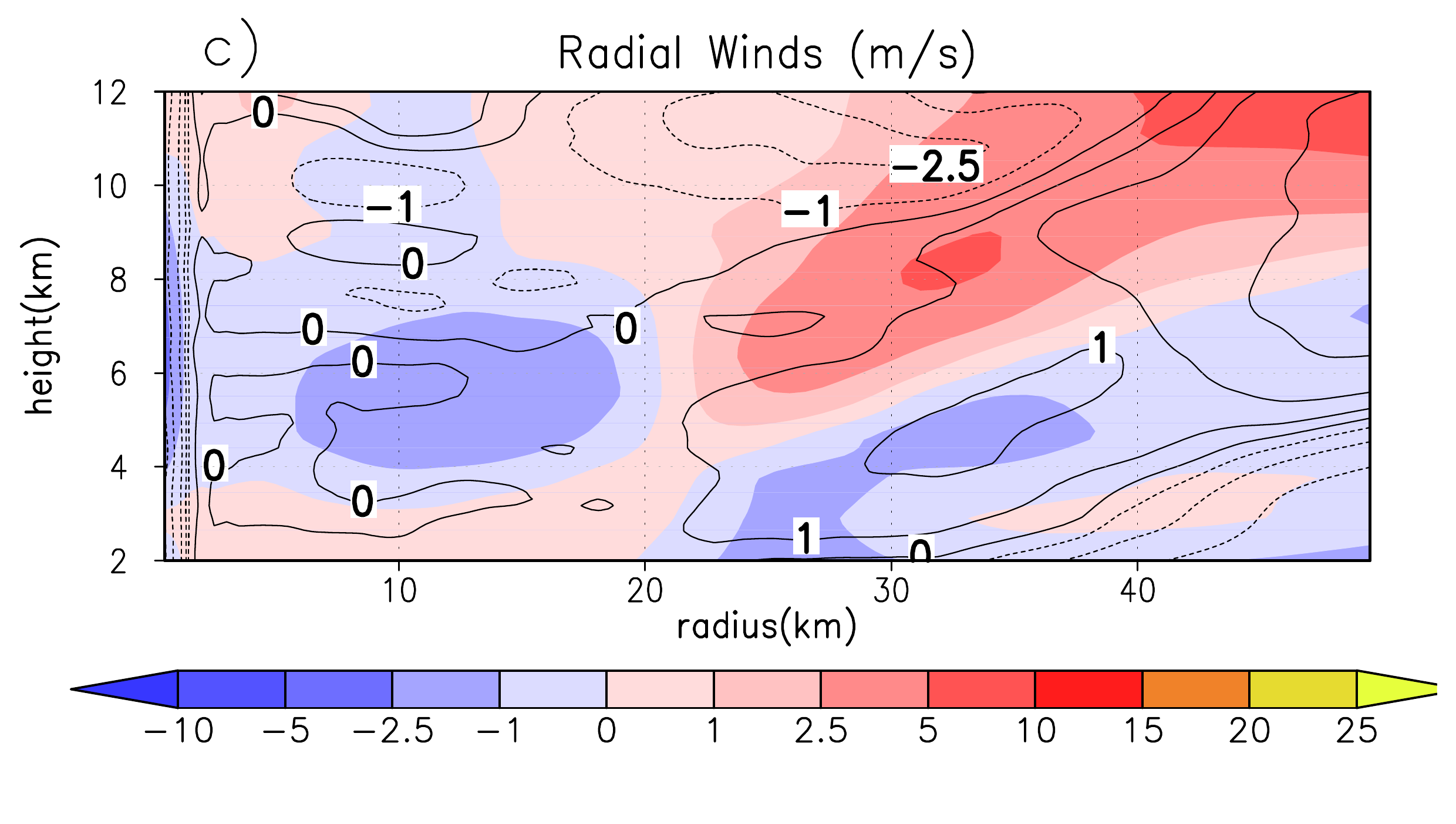}
   \includegraphics[height=0.48\linewidth,angle=-90]{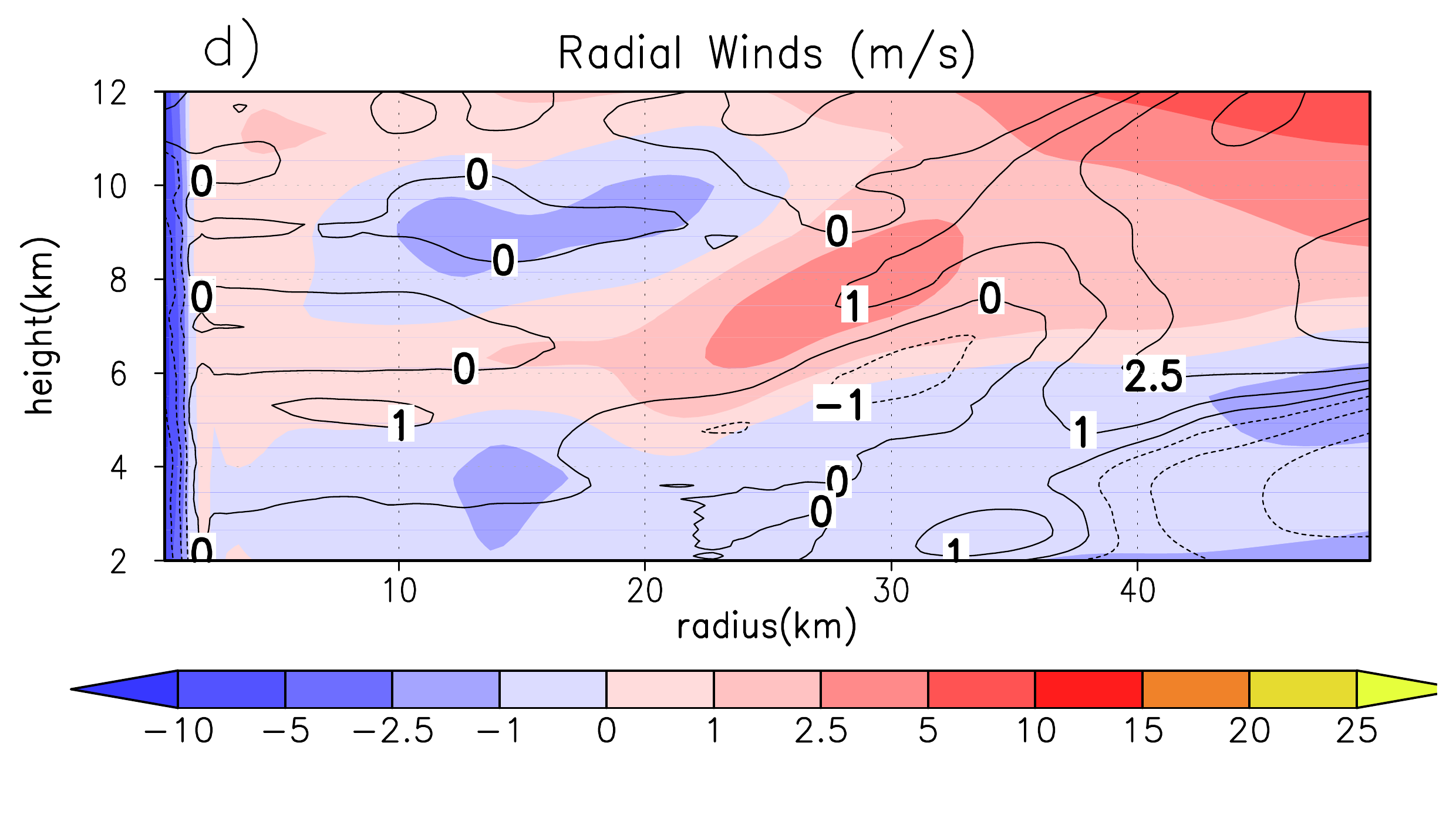}\\
   \includegraphics[height=0.48\linewidth,angle=-90]{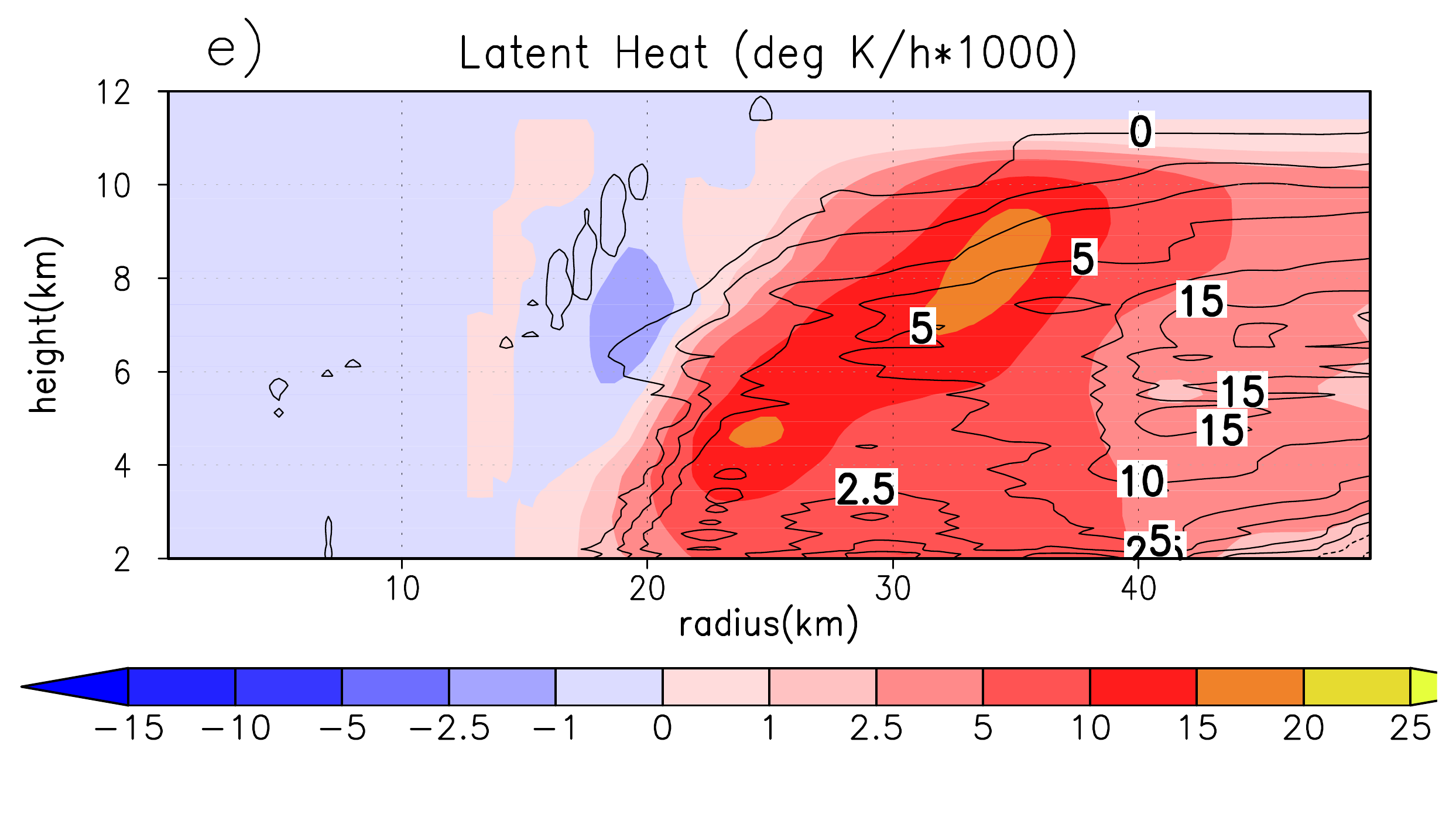}
   \includegraphics[height=0.48\linewidth,angle=-90]{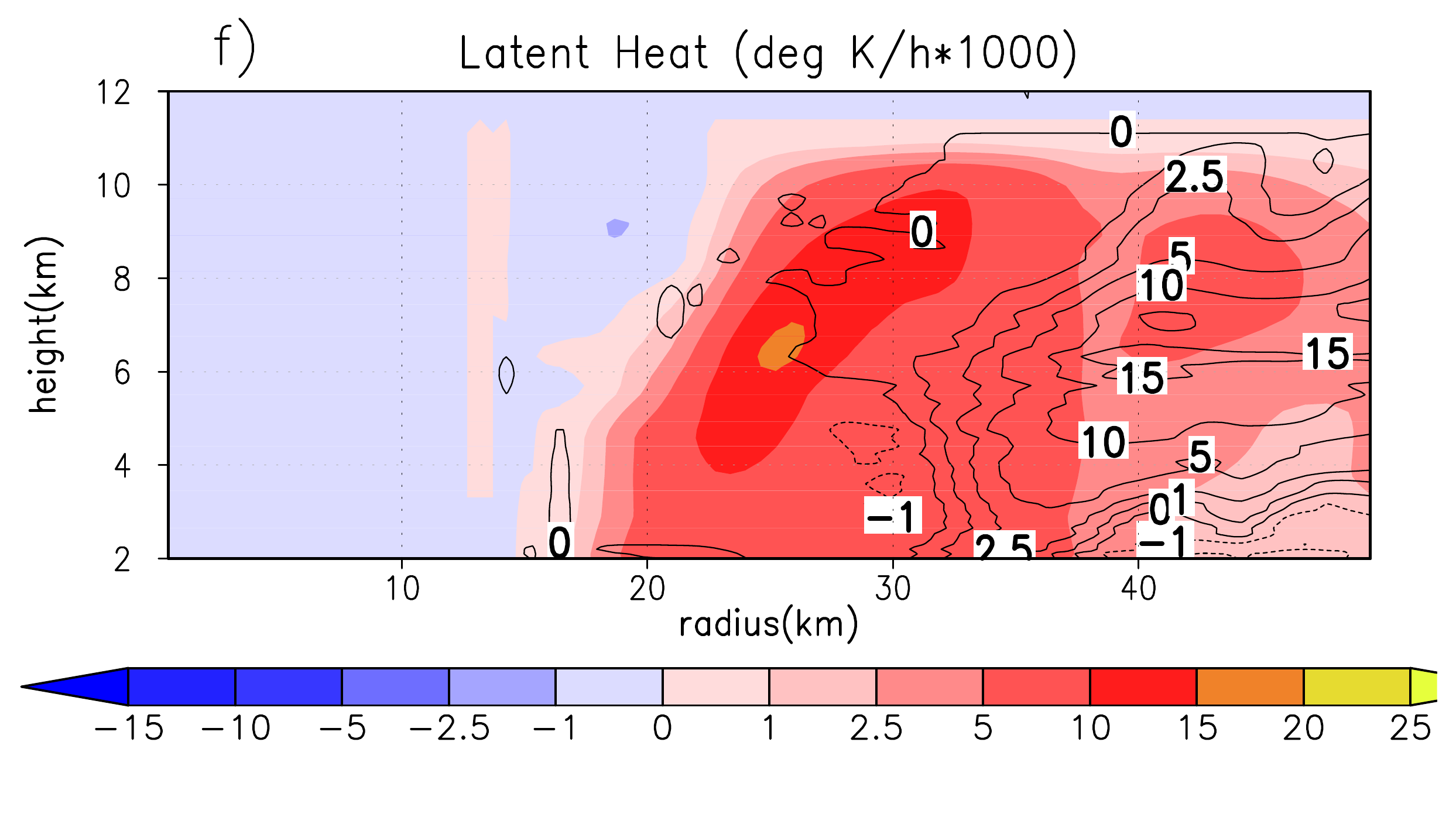}
   \caption{ Comparisons of the azimuthally-averaged profiles between
   a model simulation using estimated parameters
   from DA2 and observations (color shaded). Plots for
   tangential winds (top), radial winds (center), and latent heat (bottom)
   for the same time periods as in the previous figure.}
   \label{fig:ModelObsUV}
\end{figure}
\begin{figure}
   \centering
   \includegraphics[width=\linewidth]{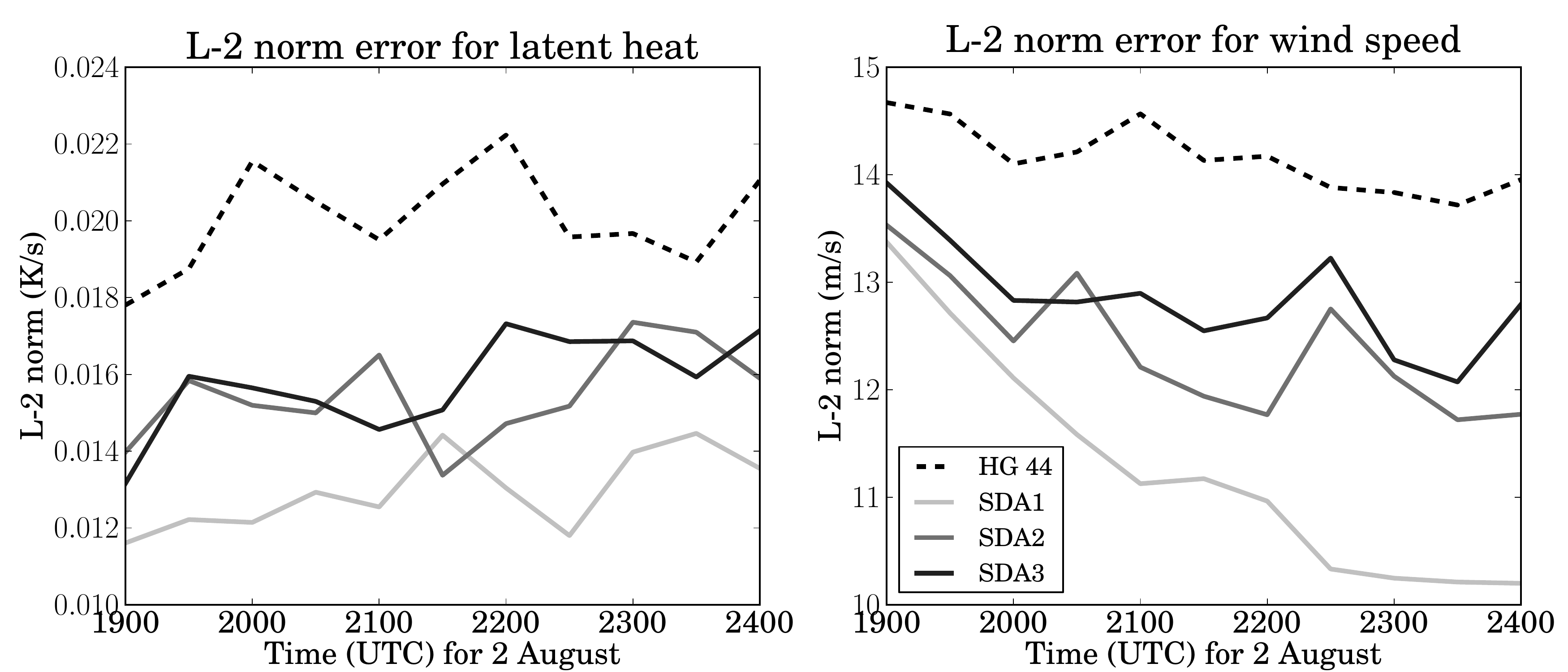}
   \caption{Error estimates as a function of time computed using Eq.
   20 of \cite{ReisnerJeffery2009} for ensemble member 44 (HG 44), SDA1,
   SDA2, and SDA3.}
   \label{l2error}
\end{figure}

To illustrate the ability of the parameter estimates to reduce model
forecast error, three simulations (SDA1, SDA2, and SDA3) were run
using the time average parameter estimates from DA1, DA2, and DA3
shown in table~\ref{tab:ParameterTimeAverage}. The setup for each
simulation is the same as described in
Subsection~\ref{sec:ModelSetupStructure}.  Both qualitatively (see
Figs.~\ref{fig:ModelObsLH}-\ref{fig:ModelObsUV}) and quantitatively
(see Fig.~\ref{l2error}) the parameter estimates produce model fields
that are in better agreement with observed fields, especially the
estimates associated with DA1 or those derived from using the latent
heat fields. Though it is not surprising that SDA1 best matches the
observed latent heat fields, what was surprising was the errors
associated with the wind fields were lower in SDA1 than those
associated with SDA2. This suggests the possible utility of using
latent heat instead of more traditional observational fields such as
horizontal winds or radar reflectivity within data assimilation
procedures to estimate model parameters and/or portions of the model
state vector. When a combination of both observational fields are
utilized for parameter estimation, error estimates from SDA3 reveal
that the response is weighted towards producing results closer to
SDA2, suggesting the dominance of the horizontal wind observational
data in the parameter estimates. This behavior can also be appreciated
in the parameter estimates themselves, as seen in
Figs.~\ref{fig:AnalysisParamterTime}~and~\ref{fig:AnalysisParameterAverage}.

\begin{figure}
   \centering
   \includegraphics[width=0.7\linewidth]{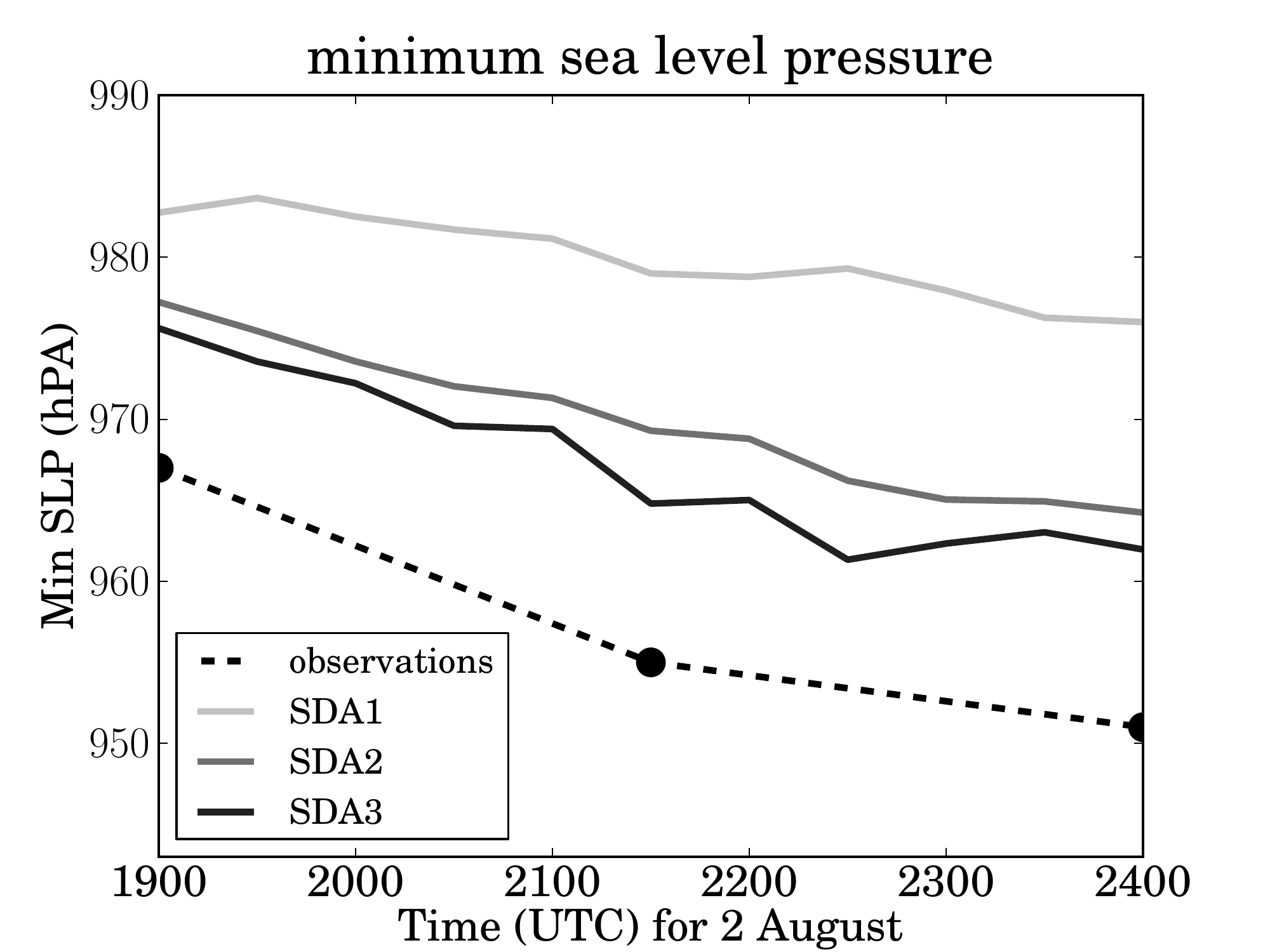}
   \caption{Minimum sea level pressure for SDA1, SDA2, and SDA3 along
   with the observed pressure from Hurricane Guillermo.}
   \label{fig:SLP_DA1DA2DA3}
\end{figure}
Though Fig.~\ref{l2error} demonstrates that SDA1 produces lower
errors than SDA2 and SDA3, Fig.~\ref{fig:SLP_DA1DA2DA3} illustrates
that the intensity of SDA1 is significantly weaker than SDA3 and
slightly weaker than SDA2. This finding may suggest that in
order for SDA1 to accurately reproduce the intensity of Guillermo
additional observational data is needed below the range of the radar
(approximately 1 km in height) or that other errors, such as numerical
errors, are contributing to the intensity differences, i.e., numerical
diffusion and spurious evaporation associated with large numerical
errors found near cloud boundaries.  Note, like ensemble member 44,
SDA2 produces nearly twice the observed amount of latent heat (see
Fig.~\ref{fig:ModelObsUV}) suggesting the simulation is indeed having
to compensate for large amounts of spurious evaporative cooling.

For example, the bottom panels of
Figs.~\ref{fig:ModelObsLH}-\ref{fig:ModelObsUV} show areas of
evaporative cooling occurring immediately to the left of regions of
strong positive latent heat release that do not have an analog in the
observed fields. Hence, similar to what is shown in Fig. 3b of
\cite{margolinvof}, this spurious cooling appears to be the result of
not being able to resolve the sub-grid movement of cloud boundaries,
i.e., the so-called advection-condensation problem.  In fact, when a
simulation using the evaporative limiter described in
\cite{ReisnerJeffery2009} along with the parameter values from DA1 is run
the resulting minimum sea level pressure from this simulation is
actually slightly lower than the observed pressure (not shown). But,
though SDA1 appears to suffer from relatively large numerical errors
that are also common to all hurricane models, the end result of the
parameter estimation procedure is still a reduction in overall model
forecast error.

\section{Summary and conclusions}
\label{sec:SummaryConclusions}

This paper presented the estimation of key model parameters found
within a hurricane model, through the use of EnKF data assimilation.
The particular approach taken was to use the EnKF to only estimate the
parameters at each time instance where observations were available.
The advantage of this approach is that it avoids the combination of
dynamic and non-dynamic elements in the assimilation procedure, which
introduces difficulties when estimating parameters.
An efficient matrix-free EnKF data assimilation algorithm
\cite{GodinezMoulton2012} is used to assimilate the derived data
fields; namely horizontal wind or latent heat, available for Hurricane
Guillermo.  Likewise, upon completion of a 120 member ensemble that
reasonably reproduced observations, the parameter estimation
experiments show that a large number of data points are indeed
required within the current approach to provide a reasonable estimate
of the model parameters. Nevertheless, the parameter estimation
procedure presented in this work can be easily be applied to other
models and data sets.

A unique aspect of this work was the utilization of derived fields of
latent heat to estimate the parameters. The estimates obtained using
these derived fields produced lower model forecast errors than a
simulation using parameter estimates obtained from horizontal wind
fields or radar reflectivity alone (not shown).
Unlike latent heat which can be directly linked to a simple physical
process occurring within a hurricane model, i.e., condensation,
utilization of other data fields such as radar reflectivity require
the model to faithfully capture physical processes that are not yet
well understood, i.e.  collision-coalescence; and are also not the
primary driver for hurricane intensification, potentially leading to
large errors in parameter estimates.  This result also suggests the
parameters associated with the primary component of hurricane
intensification, condensation of water vapor into cloud water, should
also be included in the current parameter estimation procedure.
%
It is important to note that deriving latent heat fields requires
accurate vertical velocity measurements, which in most cases are not
available. The availability of dual-Doppler radar data, for Hurricane
Guillermo, made the computation of latent heat possible. Such a data
set might not be easily acquired for other hurricanes, but one of the
contributions of this work is to demonstrate the value of such a data
set for parameter estimation.

Another subtle aspect suggested by this paper is that in order for a
given hurricane model to both reproduce a realistic latent heat field
and the correct intensity, numerical errors, especially near cloud
edges, must be small. Currently, all hurricane models produce large
numerical errors near cloud boundaries with these errors possibly
inducing significant amounts of spurious evaporation. Hence, future
work is needed to help reduce the impact of cloud-edge errors either
via the calibration of a tuning coefficient employed within an
evaporative limiter, i.e., see Eq.~A24 in \cite{ReisnerJeffery2009}, using
the current EnKF procedure or replacing Eulerian cloud modeling
approaches with a potentially more accurate Lagrangian approach
(\cite{mirek1}). 

\noindent {\it Acknowledgments.}
This work was supported by the Laboratory Directed Research and
Development Program of Los Alamos National Laboratory, which is under
the auspices of the National Nuclear Security Administration of the
U.S. Department of Energy under DOE Contracts W-7405-ENG-36 and
LA-UR-10-04291.  Computer resources were provided both by the
Computing Division at Los Alamos and the Oak Ridge National Laboratory
Cray clusters. Approved for public release, LA-UR-11-10121.

\begin{appendix}
\section*{\begin{center}Appendix A: EnKF equations for parameter
   estimation \end{center}}

Many studies have used the EnKF data assimilation to simultaneously
estimate model state and parameters. This can be achieved by using an
augmented state vector where the parameters are appended at the end of
the vector. In this appendix we review the EnKF equations for the
simultaneous state and parameter estimation and extract the necessary
equations for parameter estimation.

Let $\mathbf{p} \in \mathbb{R}^{\ell}$ be a vector holding the
model parameters, and $\mathbf{x}^{f} \in \mathbb{R}^{n}$ be
the model state forecast. Define the augmented state vector
\begin{equation}
   \mathbf{w} = 
   \begin{bmatrix}
      \mathbf{x}^{f}\\
      \mathbf{p}
   \end{bmatrix} \in \mathbb{R}^{n + \ell}
   \label{eq:augmentedstate}
\end{equation}
and let $\mathbf{w}_{i}$ for $i=1\ldots N$ be an ensemble of model
state forecast and parameters. For a vector of $m$ observations
$\mathbf{y}^{o} \in \mathbb{R}^{m}$ the EnKF analysis equations are
given by
\begin{eqnarray}
   \mathbf{w}^{a}_{i} &=& \mathbf{w}_{i} + \tilde{\mathbf{K}}\left(
   \mathbf{y}^{o}_{i} - \mathbf{H} \mathbf{x}^{f}_{i} \right), \qquad
   i=1,\ldots,N \label{eq:EnKF1-augmented}\\
   \tilde{\mathbf{K}} &=& 
   \begin{bmatrix}
      \mathbf{P}^{f} & \mathbf{C} \\
      \mathbf{C}^{T} & \mathbf{B}
   \end{bmatrix}
   \begin{bmatrix}
      \mathbf{H}^{T} \\
      \mathbf{0}
   \end{bmatrix}
   \left( \mathbf{H} \mathbf{P}^{f} \mathbf{H}^{T} + \mathbf{R}
   \right)^{-1},
   \label{eq:EnKF2-augmented}
\end{eqnarray}
where $\mathbf{P}^{f} \in \mathbb{R}^{n \times n}$ is the model
forecast covariance matrix, $\mathbf{B} \in \mathbb{R}^{\ell \times
\ell}$ is the parameter covariance matrix, $\mathbf{C} \in
\mathbb{R}^{n \times \ell}$ is the cross-correlation matrix between
the model forecast and parameters, $\mathbf{R} \in \mathbb{R}^{m
\times m}$ is the observations covariance matrix, $\mathbf{H} \in
\mathbb{R}^{m \times n}$ is an observation operator matrix that maps
state variables onto observations, and $\mathbf{y}^{o}_{i}$ is a
perturbed observation vector. The parameter correlation matrix is
given by
\begin{equation}
   \mathbf{B} = 
   \frac{1}{N-1} \sum^{N}_{i=1} \left(
   \mathbf{p}_{i} - \bar{\mathbf{p}} \right)\left(
   \mathbf{p}_{i} - \bar{\mathbf{p}} \right)^{T},
   \label{eq:ParameterCorr}
\end{equation}
and the cross-correlation matrix is give by
\begin{equation}
   \mathbf{C} = 
   \frac{1}{N-1} \sum^{N}_{i=1} \left(
   \mathbf{x}^{f}_{i} - \bar{\mathbf{x}}^{f} \right)\left(
   \mathbf{p}_{i} - \bar{\mathbf{p}} \right)^{T},
   \label{eq:CrossCorr}
\end{equation}
where $\bar{\mathbf{x}}^{f}$ and $\bar{\mathbf{p}}$ are the ensemble
average of the model forecast and parameters, respectively.

The system of equations
\eqref{eq:EnKF1-augmented}-\eqref{eq:EnKF2-augmented} can be written
as
\begin{eqnarray}
   && \left( \mathbf{H}\mathbf{P}^{f}\mathbf{H}^{T} + \mathbf{R}
   \right) \mathbf{z}_{i} = \left( \mathbf{y}^{o}_{i} - \mathbf{H}
   \mathbf{x}^{f}_{i} \right) \label{eq:EnKF3-augmented} \\
   && \mathbf{w}^{a}_{i} = \mathbf{w}_{i} +
   \begin{bmatrix}
      \mathbf{P}^{f} & \mathbf{C} \\
      \mathbf{C}^{T} & \mathbf{B}
   \end{bmatrix}
   \begin{bmatrix}
      \mathbf{H}^{T} \\
      \mathbf{0}
   \end{bmatrix}
   \mathbf{z}_{i},
   \label{eq:EnKF4-augmented}
\end{eqnarray}
where the vector $\mathbf{z}_{i} \in \mathbb{R}^{m}$ is the solution
of equation \eqref{eq:EnKF3-augmented}.  The augmented matrix in
equation \eqref{eq:EnKF4-augmented} can be simplified as
\begin{linenomath*}
\begin{equation*}
   \begin{bmatrix}
      \mathbf{P}^{f} & \mathbf{C} \\
      \mathbf{C}^{T} & \mathbf{B}
   \end{bmatrix}
   \begin{bmatrix}
      \mathbf{H}^{T} \\
      \mathbf{0}
   \end{bmatrix}
   =
   \begin{bmatrix}
      \mathbf{P}^{f} \mathbf{H}^{T} \\
      \mathbf{C}^{T} \mathbf{H}^{T}
   \end{bmatrix},
\end{equation*}
\end{linenomath*}
so we have the following analysis update equations for the model
forecast and parameters:
\begin{eqnarray}
   \mathbf{x}^{a}_{i} &=& \mathbf{x}^{f}_{i} + \mathbf{P}^{f}
   \mathbf{H}^{T} \mathbf{z}_{i}, \label{eq:EnKF-forecast}\\
   \mathbf{p}^{a}_{i} &=& \mathbf{p}_{i} + \mathbf{C}^{T}
   \mathbf{H}^{T} \mathbf{z}_{i}. \label{eq:EnKF-parameters}
\end{eqnarray}

The update equation \eqref{eq:EnKF-parameters}, together with
equations \eqref{eq:CrossCorr} and \eqref{eq:EnKF3-augmented}, form a
system that estimate the model parameters for a given data set. This
is the system used in our current study.

\end{appendix}

\ifthenelse{\boolean{dc}}
{}
{\clearpage}
\bibliographystyle{ametsoc}
\bibliography{bibliography/references}

\end{document}